\documentclass[aps,prb,preprint,superscriptaddress]{revtex4-1}

\usepackage{graphicx,amssymb,bm,amsfonts,amsmath,graphics,color,epstopdf}
\usepackage[bookmarks=false,pdfstartview=FitH,colorlinks=true,citecolor=blue,linkcolor=blue]{hyperref}
\usepackage{lineno}
\newcommand{\C}{C$_{60}$}

\begin{document}
%\linenumbers

\title{Orbital character effects in the photon energy and polarization dependence of pure C$_{60}$ photoemission}

\author{Drew W. Latzke}
\thanks{D.W.L.~and C.O.-A.~contributed equally to this work.}
\affiliation{Applied Science \& Technology, University of California, Berkeley, California 94720, USA}
\affiliation{Materials Sciences Division, Lawrence Berkeley National Laboratory, Berkeley, California 94720, USA}

\author{Claudia Ojeda-Aristizabal}
\thanks{D.W.L.~and C.O.-A.~contributed equally to this work.}
\email[Corresponding author~]{Claudia.Ojeda-Aristizabal@csulb.edu}
\affiliation{Department of Physics and Astronomy, California State University, Long Beach, California 90840, USA}

\author{Jonathan D. Denlinger}
\affiliation{Advanced Light Source, Lawrence Berkeley National Laboratory, Berkeley, California 94720, USA}

\author{Ryan Reno}
\affiliation{Department of Physics and Astronomy, California State University, Long Beach, California 90840, USA}

\author{Alex Zettl}
\affiliation{Materials Sciences Division, Lawrence Berkeley National Laboratory, Berkeley, California 94720, USA}
\affiliation{Department of Physics, University of California, Berkeley, California 94720, USA}
\affiliation{Kavli Energy NanoSciences Institute at the University of California Berkeley and the Lawrence Berkeley National Laboratory, Berkeley, California 94720, USA}

\author{Alessandra Lanzara}
\email[Corresponding author~]{alanzara@lbl.gov}
\affiliation{Materials Sciences Division, Lawrence Berkeley National Laboratory, Berkeley, California 94720, USA}
\affiliation{Department of Physics, University of California, Berkeley, California 94720, USA}

\date{\today}

\begin{abstract}
Recent direct experimental observation of multiple highly-dispersive C$_{60}$ valence bands has allowed for a detailed analysis of the unique photoemission traits of these features through photon energy- and polarization-dependent measurements. Previously obscured dispersions and strong photoemission traits are now revealed by specific light polarizations. The observed intensity effects prove the locking in place of the C$_{60}$ molecules at low temperatures and the existence of an orientational order imposed by the substrate chosen. Most importantly, photon energy- and polarization-dependent effects are shown to be intimately linked with the orbital character of the C$_{60}$ band manifolds which allows for a more precise determination of the orbital character within the HOMO-2. Our observations and analysis provide important considerations for the connection between molecular and crystalline C$_{60}$ electronic structure, past and future band structure studies, and for increasingly popular C$_{60}$ electronic device applications, especially those making use of heterostructures.
%AL-Add more specific conclusions, main points, why we care
\end{abstract}

\pacs{}

\maketitle

%Begin Introduction
Fullerene C$_{60}$ has a unique buckyball molecular structure. In its crystalline form, it exhibits a number of unconventional properties promising for modern electronic device applications including photovoltaics \cite{Rao2010}, solar cells \cite{Liao2014,Liu2014}, and field-effect devices \cite{Ojeda-Aristizabal2017,Kim2015}. Interests in C$_{60}$ are broad, extending to other fields such as astrophysics, where signature of the formation of C$_{60}$ in the interstellar medium has been an active subject in recent years \cite{Cami2010,Sellgren2010,Campbell2015}. Despite its long history of study and plentiful list of promising applications, the dispersive C$_{60}$ electronic valence band structure was unable to be observed experimentally until recently \cite{Latzke2019} and its complex photon energy- and polarization-dependent photoemission effects have yet to be thoroughly investigated, especially in terms of the orbital character of crystalline C$_{60}$'s electronic states.

%Even though the band manifolds that lie closer to the Fermi energy are those that contribute most importantly to the electronic transport, optical properties or the chemical compatibility with other compounds, learning about all band manifolds that make the electronic structure of thin film C$_{60}$ can give us a wider understanding of this material, that has been shown compatible with van der Waals heterostructures.

The electronic structure of a single molecule of C$_{60}$ can be understood within a simple model \cite{Dresselhaus1996} as distorted sp$^2$ bonds (a consequence of the curved C$_{60}$ surface) confined to the shell of the molecule that couple each carbon atom to its three nearest neighbors, giving rise to three occupied bonding $\sigma$-like orbitals and one occupied bonding $\pi$-like orbitals normal to the shell.  %(see Fig.~\ref{fig:Fig2}(e)). 
In the form of a thin film (as studied here), 
%(relevant in multiple applications including van der Waals heterostructures), 
these free C$_{60}$ molecular discrete levels (either $\sigma$ or $\pi$)
%(associated to the ) 
become large band manifolds each holding a diverse orbital character opportune for investigation.

In general, knowledge about the orbital character of electronic states in a crystalline material has proven to be quite valuable as orbital character can dramatically affect photoemission data (e.g.~trigonal warping \cite{Kormanyos2013,Rostami2013}), can provide information about the dimensionality of electronic states (e.g.~two- versus three-dimensional \cite{Zhou2006a}), and may even be intimately connected  to the formation of scientifically-intriguing material phases (e.g.~antiferromagnetic and superconducting phases in Fe-based materials \cite{Thirupathaiah2010,Wang2012e}). For C$_{60}$, it has been shown that the molecular bonding with metallic substrates is highly affected by the specific orbitals involved in the interaction \cite{Lichtenberger1993} and the occupied states momentum density for solid C$_{60}$ is also highly dependent on the relevant orbital \cite{Vos1997}. For the particular case of C$_{60}$ thin films, we analyze how the orbital character of its band manifolds is linked with strong photoemission effects, which provides fundamental information about how crystalline C$_{60}$ states relate back to the corresponding molecular states.

%Better explain matrix elements--angular dependence too, not just polarization
%Expand on photon energy oscillations literature/history, also introduce any of the other polarization dependence literature

Polarization- and photon energy-dependent photoemission studies provide a useful lens to study 
%electronic chirality \cite{Park2012,Liu2011}, magnetic properties, and most importantly 
the orbital character of electronic states \cite{Dresselhaus1996}. 
Few works have studied the effects of linear- and circular-light  polarization on C$_{60}$ photoemission \cite{Schiessling2003,Fecher1997} where limited angle and energy resolution have impeded a detailed characterization of the C$_{60}$ band manifolds.
%The effect of different linear polarizations on C$_{60}$ photoemission data has remained largely unstudied. Among the few studies, Schiessling et al.~\cite{Schiessling2003} found an angular dependence of \C~photoelectron intensity where intensity is maximized when the angles of the photoelectron and linear polarization vector are equal. Similar to the aforementioned lack of linear polarization dependence literature, little has been reported on C$_{60}$'s circular polarization dependence. Fecher et al.~\cite{Fecher1997} performed circular dichroism studies on C$_{60}$ thin films drawing a main conclusion that the HOMO-2 is mainly of a $\sigma$-like character at 30~eV. While this conclusion may be true, the methods used to derive it do not hold up to our more detailed analysis. 
In contrast, there have been significant experimental and theoretical studies of the C$_{60}$ band manifold intensity oscillations with photon energy \cite{Ruedel2002,Becker2000,Frank1997,Xu1996a,Benning1991,Wu1992,Wang2008a,Toffoli2011,Colavita2001,Hasegawa1998a,Liebsch1995,Venuti1999,Decleva2001,Ton-That2003,Li2007}. There is however no reported characterization of the oscillations' phase deviations within each band manifold, in particular for the HOMO-2, which can provide information about the localization of $\pi$ and $\sigma$ states within certain energy (and momentum) ranges.

Here we report on the detailed photon energy and polarization dependence of the thin film C$_{60}$ band structure, finding signature of the different orbital characters in C$_{60}$ valence band manifolds. These observations are possible thanks to the locking of the C$_{60}$ molecules provided by the low temperature of our measurements (20~K) and the particular choice of substrate, that, as we have shown in the past \cite{Latzke2019}, imposes a constraint on the orientation of the C$_{60}$ molecules. Our linear polarization analysis reveals expansive and specific polarization-dependent intensity enhancements and patterns for the HOMO and HOMO-1 as a result of strong matrix elements effects. Completely separate bands in the HOMO-2 are observed for different linear polarizations. Additionally, we find a strong overall intensity effect for each band manifold based on orbital character. Our circular polarization analysis provides a significantly larger wealth of data, reinforces our observations of multiple bands within each band manifold, and reveals a band splitting at the top of the HOMO not resolved with linear polarizations. Finally, our photon-energy-dependent analysis of the C$_{60}$ intensity oscillations with k$_z$ determines the frequency and phase of the oscillations for the HOMO, HOMO-1, and HOMO-2 across a wide range of energy and in-plane momenta in contrast to previous studies that only looked at the overall band manifold or at a single momentum. Our results importantly reveal precise details of the differing orbital character of each band manifold especially within the HOMO-2.

%Sample/Experiment Details
High quality C$_{60}$ thin film samples were grown in situ under ultra-high vacuum on a bulk Bi$_2$Se$_3$ substrate as detailed elsewhere \cite{Latzke2019}. High-resolution ARPES experiments were performed at Beamline 4.0.3 (MERLIN) of the Advanced Light Source using 30--128~eV linearly- or circularly-polarized photons in a vacuum better than $5\times10^{-11}$~Torr. The total-energy resolution was 20~meV with an angular resolution ($\Delta\theta$) of $\leq0.2^{\circ}$. Data were taken at 20~K to assure the absence of spinning of the individual C$_{60}$ molecules, known to rotate and follow a ratcheting behavior above 50~K \cite{Dresselhaus1996}.
%End Introduction

\begin{figure}
\includegraphics[width=1.0\linewidth]{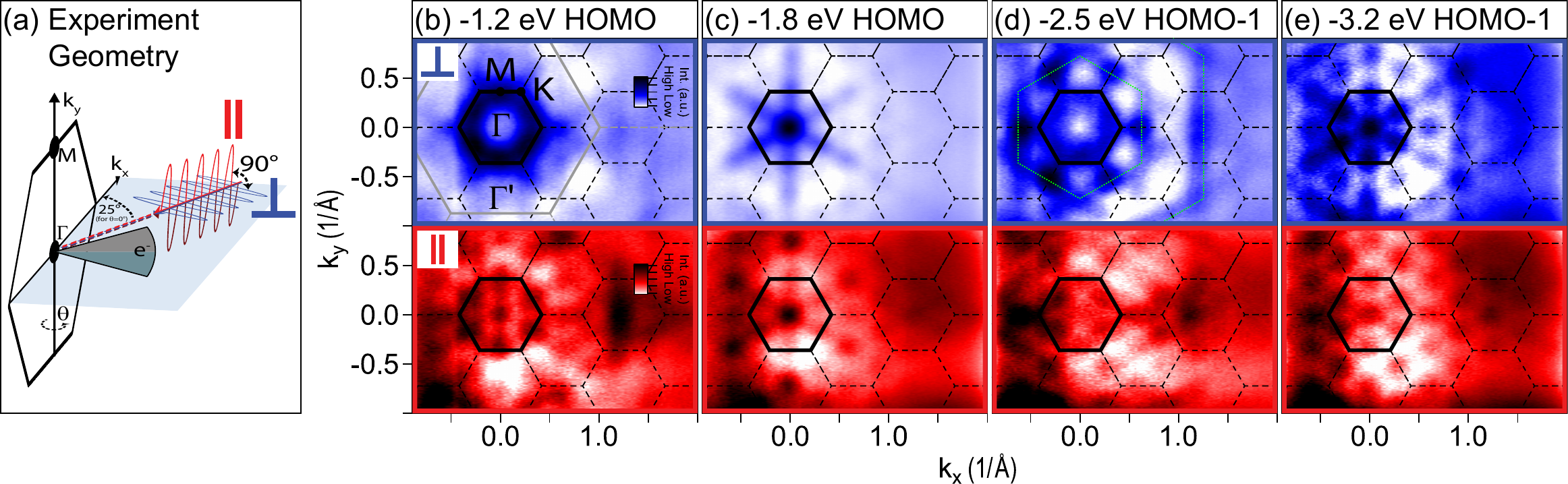}
\caption{\label{fig:Fig1}\textbf{(a)} Geometry of the experimental setup indicating incoming light polarization, outgoing photoelectrons, and Brillouin zone. High resolution constant energy maps of C$_{60}$ (5~nm) on Bi$_2$Se$_3$ with cuts through the energies near the \textbf{(b)} HOMO top, \textbf{(c)} HOMO bottom, \textbf{(d)} HOMO-1 top, and \textbf{(e)} HOMO-1 bottom. Upper and lower images are taken with out-of-plane light polarization ($\perp$, blue) and in-plane light polarization ($\parallel$, red), respectively. The first Brillouin zone of C$_{60}$ (Bi$_{2}$Se$_{3}$) is indicated by a thick black (gray) hexagon with the high-symmetry points labeled in panel (b). Higher order Brillouin zones are indicated by dashed hexagons. ($h\nu=45$~eV, $T=20$~K).}
\end{figure}

%Begin Fig. 1 Discussion
%Fig. 1: HOMO
Fig.~\ref{fig:Fig1} shows the linear-polarization-dependent band structure of thin film C$_{60}$ through ARPES constant energy maps. Panel (a) presents the experimental setup indicating sample orientation, outgoing photoelectrons, and incoming photon beam. Two orthogonal linearly-polarized beams are depicted: mostly out-of-plane light polarization ($\perp$, blue) and fully in-plane light polarization ($\parallel$, red). %(Which corresponds to a polarization parallel or perpendicular to a plane that includes the incident light and ejected photoexcited electrons respectively).  
We compare constant energy maps taken with these two orthogonal linear light polarizations at energies near the top and bottom of the HOMO and HOMO-1, as shown in panels (b--e). The rich effect of the linear light polarization observed here contrasts with previous works on a C$_{60}$ thin film \cite{Schiessling2003} that simply report an angular dependence of \C~photoelectron intensity where intensity is maximized when the angles of the photoelectron and linear polarization vector are equal. We note a strong intensity pattern following the first Brillouin zone boundary (thick black hexagon) for the top energy of the HOMO with $\perp$ polarized light (panel (b)), which is completely attenuated for $\parallel$ light. The $\parallel$ data reveals the intensity spot corresponding to the band maximum at $\Gamma$ which is entirely obscured otherwise due to the intensity effect. Similarly, we observe an enhanced intensity at $\Gamma$ for the bottom of the HOMO (panel (c)), which is reduced in the $\parallel$ case along with the asterisk-shaped intensity pattern radiating along $\Gamma-\text{M}$ into higher-order Brillouin zones (dashed-line hexagons). The intensity at the $\Gamma'$ points in these higher-order Brillouin zones is relatively increased for the $\parallel$ case and we recover a close resemblance between the first and higher-order Brillouin zones.

%Fig. 1: HOMO-1
Similar to the HOMO, the top energy of the HOMO-1 (panel (d)) has an enhanced intensity along its first Brillouin zone boundary for the $\perp$ polarization , albeit with a lesser magnitude than in the HOMO case. Additionally, we observe large strong intensity patterns outside of the first Brillouin zone marked here by dotted bright green hexagons. These patterns again nearly completely disappear for the $\parallel$ polarization. The bottom energy of the HOMO-1 (panel (e)) exhibits a slightly enhanced intensity within the first Brillouin zone and a large intensity pattern outside of it as well. Again, these effects are largely negated when the polarization is switched to $\parallel$.

Therefore, there are three important effects visible from Fig.~\ref{fig:Fig1}: (1) strong intensity enhancements and patterns that are different for the HOMO and HOMO-1, (2) a dependence of the intensity enhancements and patterns on the light polarization ($\perp$ or $\parallel$), and (3) a non-equivalence of the patterns among different Brillouin zones. Being the physical structure of the crystalline thin film periodic, one would expect corresponding periodicity and symmetry in the ARPES data which is broken by the observed intensity effects. The symmetry breaking observed here is however a well known effect for graphite and graphene. Graphite and graphene, characterized by a triangular lattice with two atoms per unit cell, are subject under photoemission experiments to a zone selection rule based on symmetry arguments. $\sigma$-like and $\pi$-like states in these systems appear with different intensity in otherwise equivalent Brillouin zones. This effect has its origin in the interference between photoelectrons emerging from the two atoms in each unit cell that follow different path lengths to the experiment's detector. At normal emission there is no difference in path lengths, while for higher-order Brillouin zones (non-normal emission) there is a non-zero difference, creating a positive and negative interference effect for the first Brillouin zone and higher-order Brillouin zones, respectively. Given that the surface of our thin film C$_{60}$ is arranged in a triangular lattice similar to graphene and graphite, a question that arises is if Effect 3 mentioned above has the same origin as the zone selection rule occurring in graphene/graphite \cite{Daimon1995,Shirley1995}.  

\begin{figure}
\includegraphics[width=0.5\linewidth]{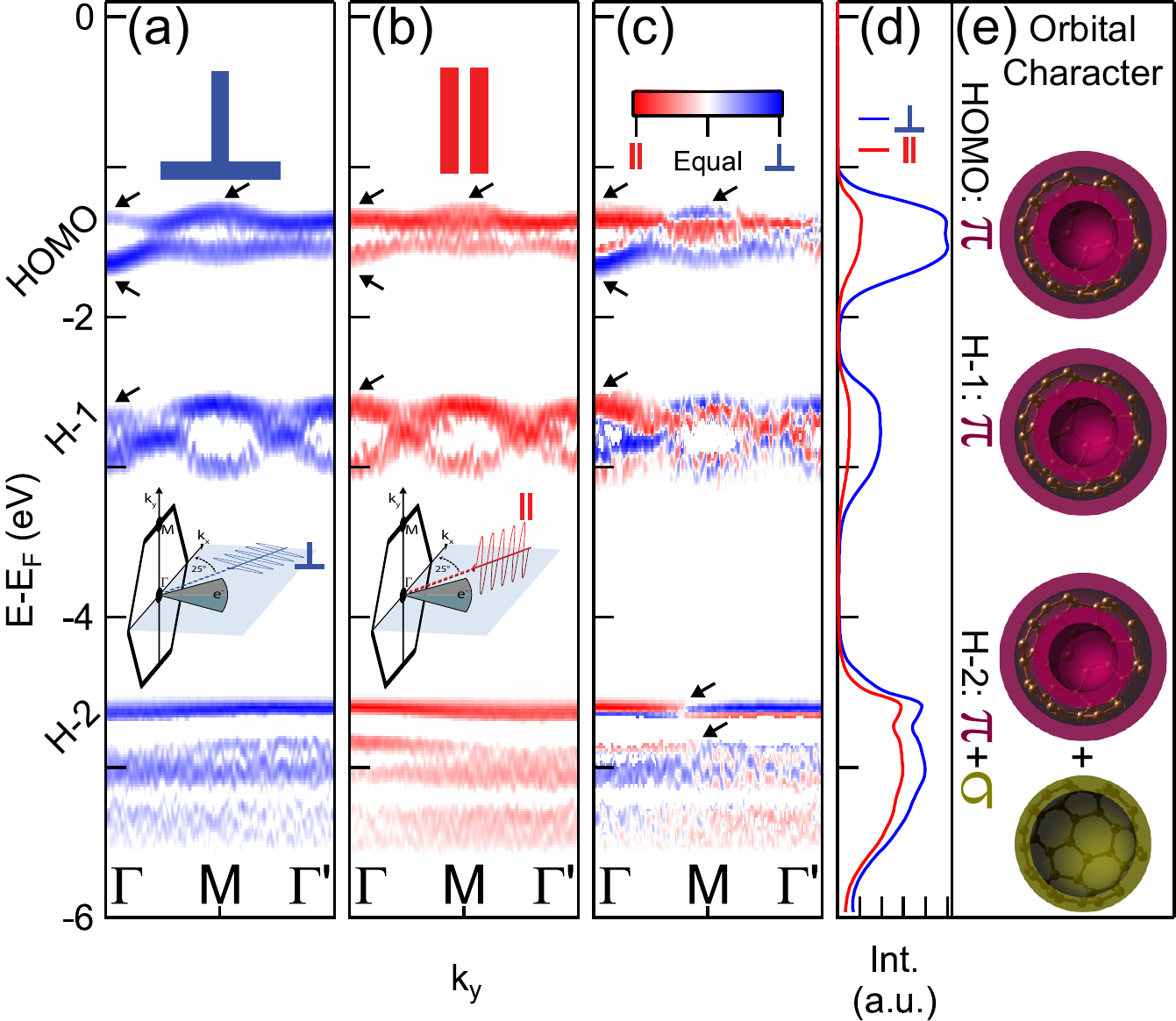}
\caption{\label{fig:Fig2}Polarization-dependent C$_{60}$ band structure along $\Gamma-\text{M}$ from ARPES curvature \cite{Zhang2011} using \textbf{(a)} out-of-plane light polarization ($\perp$, blue), \textbf{(b)} in-plane light polarization ($\parallel$, red), and \textbf{(c)} an image of their difference. See insets for polarization geometry. \textbf{(d)} Integrated (across $\Gamma'-\text{M}-\Gamma-\text{M}-\Gamma'$) EDCs for incident out-of-plane (blue) and in-plane (red) polarized light. Intensities were normalized by incident photon flux. \textbf{(e)} Cross section diagram illustrating a \C~molecule (C atoms and bonds are brown) with its theory-derived orbital character for each of the first three valence band manifolds. $\pi$-like states are localized within the ruby-colored surfaces that lie just outside and inside the buckyball surface, while $\sigma$-like states are localized within the olive-colored surface that lies along the buckyball surface.}
\end{figure}

%Fig. 1: Similar Dispersions
Elaboration on the possible cause of Effects 1 and 2 necessitates a discussion on the predicted orbital character of the HOMO and HOMO-1. Two quantum numbers can be used to label the molecular orbitals in C$_{60}$: $n$ and $l$. This labeling scheme was introduced by Martins et al.~\cite{Martins1991} who assumed that the screened electronic potential in C$_{60}$ is localized on a hollow shell encompassing the buckyball. Motivated by the spherical shape of C$_{60}$, this model takes into account the angular behavior of the molecular orbitals. The angular number $l$ describes the angular pattern of the wavefunction nodes, while the principal quantum number $n$ equals the number of radial nodes plus one. The $\sigma$ wavefunctions, which do not have a radial nodal surface as they lie on the surface on the buckyball, correspond to $n=1$, while $\pi$ orbitals having one radial nodal surface (at the buckyball surface) correspond to $n=2$. (See Fig.~\ref{fig:Fig2}(e).) Inspection of the radial functions of each molecular orbital in C$_{60}$ allows the labeling of the HOMO with $\pi_{5}$ (i.e.~$n=2$ and $l=5$) and the HOMO-1 with $\pi_{4}$ (i.e.~$n=2$ and $l=4$). In other words, the HOMO and HOMO-1 are both formed entirely from $\pi$ orbital character, but differ in angular number by one. Others have confirmed these predictions finding that the $\pi$ orbitals for both the HOMO and HOMO-1 are localized to surfaces just outside and inside the icosahedral C$_{60}$ shell \cite{Ruedel2002,Weaver1994,Kuzmany1998} with a nodal surface along the buckyball surface. The HOMO and HOMO-1 can be further distinguished as the HOMO is formed by h$_u$ orbitals, while the HOMO-1 is formed by g$_g$ and h$_g$ orbitals. Whereas these are all formed entirely by $\pi$ orbitals perpendicular to the surface at each carbon atom, the h$_g$ orbital has the particular characteristic of being 25\% in-plane as a result of renormalization due to the curved buckyball surface \cite{Lichtenberger1993}. We deduce that Effect 1, the light-polarization-dependent intensity enhancements and patterns observed in Fig.~\ref{fig:Fig1}, are due to strong photoemission matrix elements effects (most evident in the $\perp$ polarized data) which manifest in different ways for the HOMO and HOMO-1 likely as a result of their small orbital character differences described above.

In all cases we find these effects to be minimized for the $\parallel$ polarization and we are able to observe strong similarities between the HOMO and HOMO-1 dispersions as reported previously \cite{Latzke2019,He1995}. For example, when considering the $\parallel$ polarization maps, the top energy of the HOMO and HOMO-1 (bottom of panels (b) and (d), respectively) both primarily show intensity spots at $\Gamma$ and $\text{M}$ (and at equivalent locations in higher-order Brillouin zones) indicating local band maxima at these momenta. The bottom energy of the HOMO and HOMO-1 (bottom of panels (c) and (e), respectively) show similar patterns indicating local band minima at $\Gamma$ and $\text{M}$ as well. The similarity in dispersion for the HOMO and HOMO-1 is most likely a result of their overall similar $\pi$-orbital character.
%DL- cite pi orbital shell paper
%DL- discuss n, l notation earlier (here)
%End Fig. 1 Discussion 

%for pi orbital dispersion: and interact with neighboring C$_{60}$ atoms producing band dispersions. It is expected therefore that the HOMO and HOMO-1 band manifolds have similar dispersions. 
%Move the following orbital character discussion elsewhere
%$n=2$, i.e. the HOMO and HOMO-1 consist entirely of $\pi$ orbitals, while the HOMO-2 is identified as a mix of both $\pi$ and $\sigma$ states including $\pi_3$, $\sigma_8$, and $\sigma_9$ \cite{Martins1991,Benning1991} (or possibly $\pi_3$, $\sigma_9$, and $\sigma_{10}$ \cite{Colavita2001}). 
%For the HOMO and HOMO-1:They also are similar when compared with the bonding structure in dimer C$_2$ as they have predominantly $\pi$ bonding (as opposed to antibonding) character.

%Begin Fig. 2 Discussion
%Fig. 2: HOMO and HOMO-1
Fig.~\ref{fig:Fig2} compares ARPES curvature \cite{Zhang2011} in the energy dimension data along $\Gamma-\text{M}$ from each of the two orthogonal linear light polarizations. The enhanced first Brillouin zone intensity effects observed in the $\perp$ constant energy maps of Fig.~\ref{fig:Fig1}(b--e) manifest themselves as stronger peaks and/or slight shifts of the dispersions shown in Fig.~\ref{fig:Fig2}(a). For example, as seen by comparing the HOMO in the ARPES curvature image with $\perp$ polarization in panel (a) to that with $\parallel$ polarization in panel (b), the band maximum at $\Gamma$ is diminished, while the band minimum at $\Gamma$ and the band maximum at $\text{M}$ are enhanced for the $\perp$ polarization (see arrows). This is nicely summarized in panel (c) by comparing the difference of the curvature from each polarization where we can see a red color (corresponding to stronger curvature with the $\parallel$ polarization) for the HOMO band maximum at $\Gamma$, and a blue color ($\perp$ is stronger) for the HOMO band minimum at $\Gamma$ and band maximum at $\text{M}$. Similar to the HOMO, the HOMO-1 (H-1) band maximum at $\Gamma$ is also weakened for the $\perp$ polarization as shown by a strong red band maximum at $\Gamma$ in the HOMO-1 in panel (c) (see arrows). Considering only the $\parallel$ polarization for a moment, we see the same dispersions at both $\Gamma$ and $\Gamma'$ for the HOMO and HOMO-1 as one would have initially predicted when comparing equivalent momenta. Just as observed in the constant energy maps, the matrix elements intensity effects are minimized in this case, while for the $\perp$ case we see clear differences (analogous to those previously mentioned for Fig.~\ref{fig:Fig1}) when comparing $\Gamma$ and $\Gamma'$.

%Fig. 2: HOMO-2
In addition to the HOMO and HOMO-1, we are able to resolve the next highest band manifold, the HOMO-2 (H-2), as well. The polarization-dependent effects are even more striking in the HOMO-2 as the two polarizations appear to reveal, in multiple instances, completely separate bands. Starting from the top energy of the HOMO-2 for the $\perp$ polarization in Fig.~\ref{fig:Fig2}(a), we find that the first two bands have a minimum at $\Gamma$ and rise to a maximum at $\Gamma'$. However, when the polarization is switched to $\parallel$ as in panel (b), we find that the top two observable bands now have a maximum at $\Gamma$ and move down in energy to a minimum at $\Gamma'$, opposite of the behavior observed for the $\perp$ polarization. This can be observed more clearly in panel (c) as the discussed bands flip colors (for a given energy) going from $\Gamma$ to $\Gamma'$ (see arrows where flip occurs). Evidently, the effect of the polarization dependence on these top HOMO-2 bands is so strong that the two orthogonal polarizations reveal completely separate bands in the photoemission data. The HOMO-2, unlike the HOMO and HOMO-1, has majority contribution from $\sigma$-like bands which may be responsible for the specific polarization-dependent bands pointed out here as we do not see such features in the entirely $\pi$-like HOMO and HOMO-1. A possible explanation for Effect 2, the differences observed between $\perp$ and $\parallel$ polarizations for all of the band manifolds, may be related to the highly two-dimensional nature of the C$_{60}$ thin film as it is only $\sim$5 layers thick while it simultaneously exhibits high lateral order. These characteristics differentiate the in-plane and out-of-plane physical structure which may each couple differently with the in-plane and out-of-plane light polarizations.
%Previous studies could not take this into account

%Fig. 2: Integrated EDC
Fig.~\ref{fig:Fig2}(d) shows a momentum-integrated energy distribution curve (EDC) comparison for the $\perp$ (blue) and $\parallel$ (red) polarizations which are normalized by incident photon flux. The overall intensity within the HOMO and HOMO-1 band manifolds for the $\parallel$ polarization is greatly reduced by 79\% and 72\%, respectively, as compared with the $\perp$ polarization. However, the HOMO-2 intensity is hardly diminished comparatively as it is reduced by only 24\%. We believe this to be a consequence of the differing overall orbital character of the HOMO and HOMO-1 compared with the HOMO-2. As previously discussed, the HOMO and HOMO-1 are of entirely $\pi$ orbital character. However, the HOMO-2 is a mixture of $\pi$ (g$_u$ and t$_{2u}$ orbitals) and $\sigma$ (h$_u$ and h$_g$ orbitals) states, with more than half of the HOMO-2 electrons being of a $\sigma$ nature \cite{Troullier1992,Haddon1992}. As shown in panel (e), the physical implications of this are that the $\pi$-like states are localized to a surface outside and inside the C$_{60}$ shell with a nodal surface along the buckyball surface (as is the case for the HOMO, HOMO-1, and partially the HOMO-2), while the $\sigma$-like states are localized to a surface on the C$_{60}$ shell (as is the case mostly for the HOMO-2). In the HOMO, HOMO-1, and HOMO-2, the photoemission signal is evidently minimized from the $\pi$-like states for the case of $\parallel$ incident light, while the $\sigma$-like states are left mostly unattenuated.

%Fig. 2: Asymmetry Parameter
A more in-depth explanation of the linear-polarization effect on overall band manifold intensity may be found by considering the available experimental data and theory regarding the photon energy and polarization-dependent partial cross sections of the band manifolds. The partial differential cross sections have been calculated to be $d\sigma /d\Omega =({\sigma}_{tot}/4\pi )[1+\beta P_2(\cos\phi)]$ considering electric dipole absorption with a single-electron model \cite{Cooper1969,He2007,Cooper1968} (for total cross section ${\sigma}_{tot}$, asymmetry parameter $\beta$, second order Legendre polynomial $P_2$, and angle between outgoing photoelectron and incoming polarization vector $\phi$). The asymmetry parameter $\beta$ is dependent on the orbital character of each band manifold, particularly the angular momentum quantum number $l$. It is also dependent on photon energy and for 45~eV, it is found to be 0.95, 1.1, and 0.5 for the HOMO, HOMO-1, and HOMO-2, respectively \cite{Venuti1999,Liebsch1995}. Given $\phi$'s dependence on the polarization vector, when switching from $\perp$ to $\parallel$ the partial cross section would decrease by approximately 66\% and 71\% for the HOMO and HOMO-1, respectively, but only 42\% for the HOMO-2. A decrease in partial cross section would lead to a proportional decrease in observed intensity. Hence, the large decrease of the HOMO and HOMO-1 partial cross sections and the smaller decrease of the HOMO-2 partial cross section, while not an exact quantitative match, is consistent with our observations and likely a contributing factor to the polarization-dependent overall band manifold intensity effect observed in addition to the $\pi$ and $\sigma$ orbital character effect previously discussed. In either case, the differences in the band manifolds' orbital character are the driving forces behind this effect.
%End Fig. 2 Discussion

\begin{figure}
\includegraphics[width=0.5\linewidth]{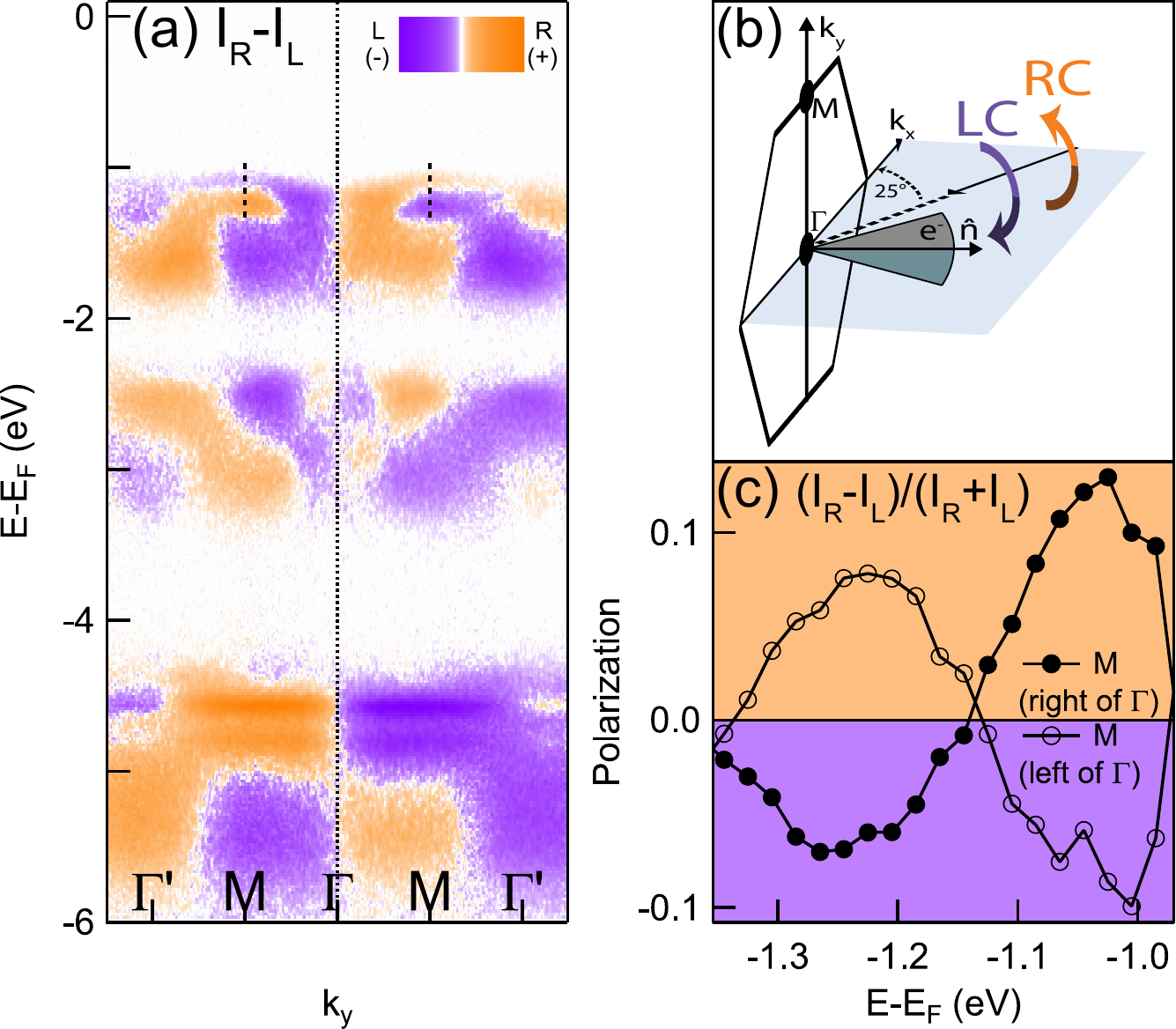}
\caption{\label{fig:Fig3}\textbf{(a)} Circular dichroism intensity difference ($\text{I}_{\text{R}}-\text{I}_{\text{L}}$) of C$_{60}$ band structure along $\Gamma-\text{M}$ where orange (purple) indicates a stronger intensity from right-hand (left-hand) circularly polarized incident light. \textbf{(b)} Geometry of the experimental setup indicating incoming right-hand (RC, orange) or left-hand (LC, purple) circularly polarized light, outgoing photoelectrons, sample normal \^n, and Brillouin zone. \textbf{(c)} Circular dichroism polarization ($[\text{I}_{\text{R}}-\text{I}_{\text{L}}]/[\text{I}_{\text{R}}+\text{I}_{\text{L}}]$) versus energy near the top energy of the HOMO at the $\text{M}$ points on either side of $\Gamma$ (dashed lines in panel (a)).}
\end{figure}

%Begin Fig. 3 Discussion
In order to fully explore the nature and orbital character details of C$_{60}$'s valence bands, we continued our polarization-dependence study by measuring the corresponding ARPES circular dichroism (CD) along $\Gamma-\text{M}$ spanning the range of two Brillouin zones, as shown in Fig.~\ref{fig:Fig3}, by using left- and right-hand circularly polarized light. CD effects are known to occur in all spatially oriented molecular thin films when using an experimental geometry with a handedness \cite{Westphal1990,Schoenhense1990}. This handedness is achieved in our experiment (panel (b)) when the direction of the incoming left-hand (LC) or right-hand (RC) circularly polarized photon beam, the sample normal \^n, and the outgoing photoelectrons ($\mathrm{e^-}$) do not lie in the same plane as is the case for all displayed momenta in panel (a) except at precisely $\Gamma$ (coincident with \^n) where the circular dichroism effect disappears as a result of the loss of handedness. The handedness combined with the symmetry of the C$_{60}$ thin film is the cause of the change in sign of the CD intensity when crossing $\Gamma$.

Using circularly-polarized light, we resolve similar dispersions as observed with linearly-polarized light, particularly for the HOMO-1 where in Fig.~\ref{fig:Fig3}(a) we see two main band groupings near the top and bottom energies differentiated by opposing signs of circular dichroism intensity ($\text{I}_{\text{R}}-\text{I}_{\text{L}}$) indicated by the orange (positive, $\text{I}_{\text{R}}>\text{I}_{\text{L}}$) and purple (negative, $\text{I}_{\text{L}}>\text{I}_{\text{R}}$) colors. The HOMO has a similar structure, but the upper energy grouping is further split in two to reveal separate features that were previously combined for either of the linear polarizations. It is only in our CD data that this more detailed structure is observed. As for the HOMO-2, similar to that observed with linear polarizations, we resolve two weakly dispersing features between -4.43~eV and -4.95~eV with similar CD polarizations which are opposite in sign of the CD polarization of the bottom half (-4.95~eV to -5.8~eV) of the HOMO-2. These HOMO-2 features will be discussed later on with regards to our photon-energy-dependent observations. Overall, we see the same structure in our circular dichroism on either side of $\Gamma$ demonstrating a proper normalization of the incident photon flux and promoting the validity of the circular dichroism features we observe.

A previous CD study on C$_{60}$ thin films \cite{Fecher1997} found that, when compared with one another, the HOMO and HOMO-1 showed the same sign and a similar magnitude of CD polarization while the HOMO-2 showed an opposite sign of CD polarization. They concluded that the opposing polarization of the HOMO-2 must originate from its (mostly) $\sigma$ orbital character based on the fact that opposing signs of CD were observed for $\sigma$ and $\pi$ bands in graphite \cite{Schoenhense1992}. This reasoning does not hold up to our more detailed analysis as we observe opposing signs of CD within even just the HOMO (or HOMO-1), which has entirely $\pi$ orbital character (as does the HOMO-1). Therefore, we cannot conclude the overall band manifold character from the overall CD for each band manifold, but this does not preclude observing more precise orbital character effects from the bands within each band manifold.

We more closely examine the polarization change across the upper HOMO splitting at the $\text{M}$ point on either side of $\Gamma$ in Fig.~\ref{fig:Fig3}(c). The curves with filled and open circles correspond to the polarization at $\text{M}$ on the right and left side of $\Gamma$, respectively. The CD signal is nearly perfectly inverted on either side of $\Gamma$, as expected. We see substantial CD polarization in this split feature as the polarization of the upper feature reaches a magnitude of approximately 0.10, while the lower feature reaches a magnitude of approximately 0.08 (with opposite sign). The strong CD polarization observed here is likely favored by the highly ordered C$_{60}$ thin film, a result of epitaxial growth on and interaction with Bi$_2$Se$_3$ as detailed elsewhere \cite{Latzke2019}. 
Similarly, previous reports on molecular thin films \cite{Westphal1990} have observed a decreased CD polarization with increasing disorder. We note that the strength of CD polarization may also depend on the angle of incidence of the incoming light (angle shown in panel (b)). For example, a sine factor based on the angle of incoming light is added to the CD polarization in the case of a simple p$_\text{z}$ orbital \cite{Dubs1985,Schoenhense1990}. C$_{60}$, however, does not have such simply-oriented orbitals (e.g.~a p$_\text{z}$ orbital) given its unique buckyball shape so is unlikely to have such a strong dependence on the angle of incoming light, but there may be some geometric effect given its two-dimensional thin film nature as opposed to that of a fully three-dimensional bulk sample.
%DL-could perhaps integrate metal bonding to C$_{60}$ paper when discussing C$_{60}$/BiSe
%CD reveals orbital character differences of these features
%Schoenhense1990:CD reveals the transition matrix elements and relative phases of the various outgoing partial waves.
%Scho: relative phases of the final-state partial waves. For atoms, these depend on l only, whereas for band states in solids (as for molecules) they depend on l and m.
%Fecher: They see no temp. dep. in CD down to 90K => no structural change at the surface.
%End Fig. 3 Discussion

\begin{figure}
\includegraphics[width=1.0\linewidth]{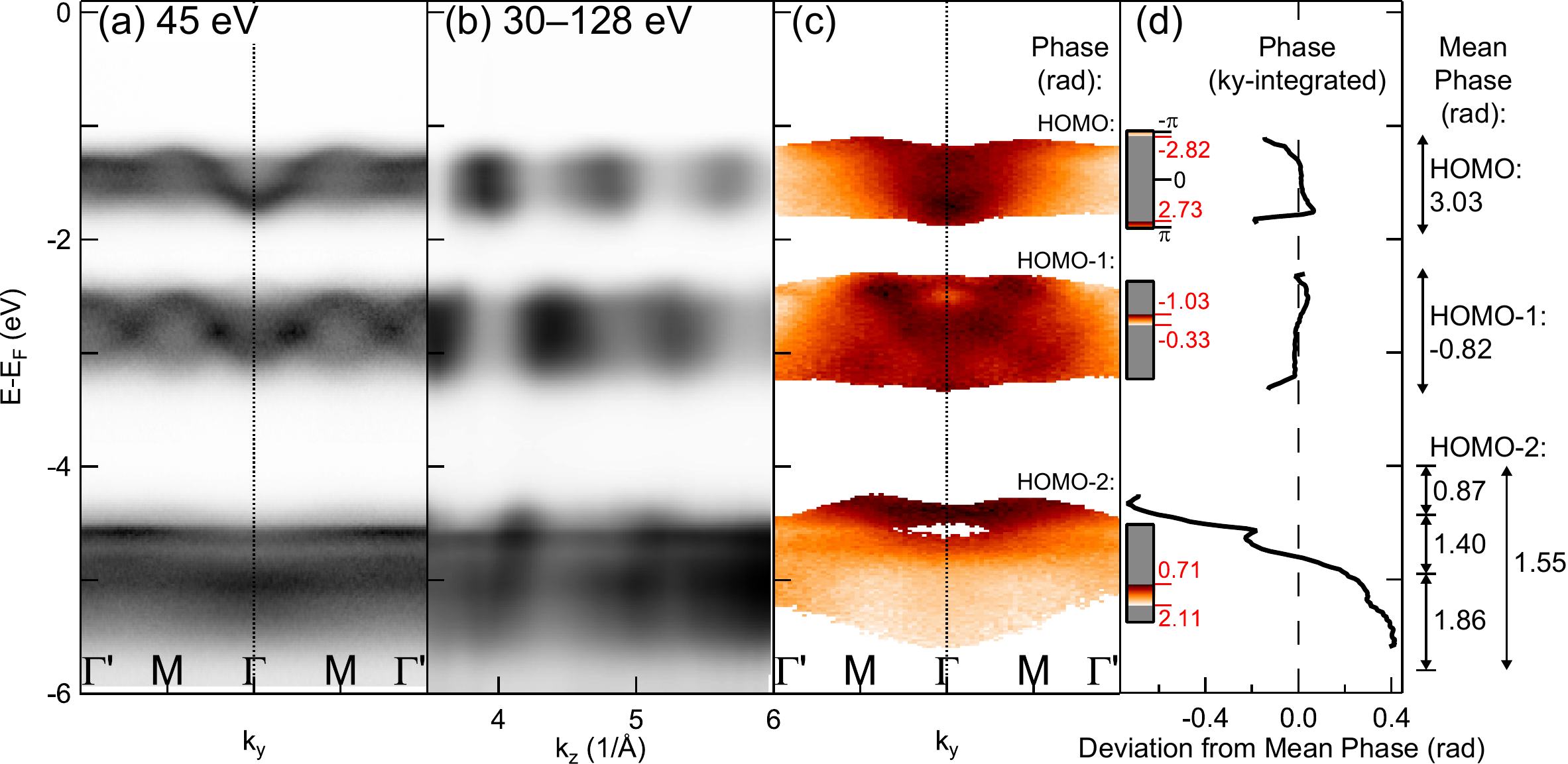}
\caption{\label{fig:Fig4}\textbf{(a)} High-resolution ARPES band structure of C$_{60}$ along $\Gamma-\text{M}$ at 45~eV. \textbf{(b)} Integrated ARPES intensity across the momentum range in panel (a) for 30--128~eV showing intensity oscillations with respect to k$_z$ in the HOMO, HOMO-1, and HOMO-2. \textbf{(c)} Extracted relative phase of the oscillations for each in-plane momentum (k$_y$) and energy shown (note the separate color scale limits for each band manifold to the right of the panel). \textbf{(d)} Deviation of phase (momentum-integrated across full k$_y$ window) from mean phase for each band manifold. Mean phase for each band manifold shown to the right (including subdivided energy ranges within the HOMO-2). All listed phase values are with regards to the range $\{-\pi, \pi$\}.}
\end{figure}

%Begin Fig. 4 Discussion
%Photon energy dependence below
In Fig.~\ref{fig:Fig4}, we present our findings on the strong oscillations of the band manifolds' intensity with photon energy (and hence k$_z$). Similar oscillations have been reported in previous works on C$_{60}$ bulk crystals \cite{Wu1992,He2007} and thin films \cite{Hasegawa1998,Wang2008a,Ton-That2003} (See Supp. Mat. Fig. S2 for a comparison with our data). High resolution data was taken for the HOMO, HOMO-1, and HOMO-2 bands along $\Gamma-\text{M}$ spanning a range of two Brillouin zones across a photon energy range of 30--128~eV (step size of 2~eV) with  $\perp$-polarized photons. We have found that the light polarization has no important effect on the intensity oscillations besides a larger peak of intensity at around 45eV for $\perp$-polarized light (See Fig. S1 Supp. Mat.). In Fig.~\ref{fig:Fig4}(a) we present data at 45~eV incident photon energy across the range of binding energy and in-plane momenta (k$_y$) considered in our oscillation analysis. Strong oscillations of the band manifolds' intensity with k$_z$ can be observed in Fig.~\ref{fig:Fig4}(b) where intensities were obtained by momentum-integrating across the k$_y$ range in panel (a). In the past, these oscillations were thought to be related to strong final state effects \cite{Wu1992}, but were later identified as a signature of interference of photoelectron waves emanating from each of the constituent C$_{60}$ carbon atoms, where each wave has a different phase. The unique spherical structure of C$_{60}$ and its large radius play a critical role in this effect \cite{Hasegawa1998,Becker2000,Ruedel2002,Ton-That2003}.

To further investigate the observed k$_z$-dependent oscillations and their link to the orbital character of the different band manifolds, we performed Fourier analysis using a Fast Fourier Transform (FFT) algorithm (see Supplemental Material). This analysis was used to separate the intensity oscillation (with respect to k$_z$) curves into different frequency components. We were able to determine the dominant frequency of the oscillation and its corresponding phase. The dominant frequency (as shown in the Supplemental Material) of the oscillations is nearly identical in each band manifold ($\approx$7.5~\AA), which is very close to the molecule's structural diameter of 7.1~\AA~\cite{Satpathy1992,PhotENote,Ruedel2002}, as expected. 
%DL: as can be expected based on the theoretical model of the oscillations
In contrast, the phase of the oscillations, as shown in Figure~\ref{fig:Fig4}(c) and (d), is quite different for each band manifold (see individual color scale limits in the range $\{-\pi, \pi$\} to the right of panel (c)). This phase dissimilarity has its origin in an interference effect, and has been explained in the past \cite{Hasegawa1998,Hasegawa1998a} after calculations of the differential photoionization cross section that approximate the molecular orbital initial state to spherical Bessel functions $j_l(kR)$ that depend on the radius of the molecule $R$, the photon-energy-dependent electron momentum $k$, and the angular quantum number $l$, that together with the principal quantum number $n$, label the molecular orbitals in C$_{60}$  as previously described.

The angular quantum number for the initial $\pi$ HOMO, HOMO-1, and HOMO-2 states can be identified as $l=5$, $l=4$, and $l=3$, respectively \cite{Martins1991}, and the final states are determined by the dipole selection rules, e.g. $l_i\Rightarrow l_f=l_i\pm1$. Mathematically, the spherical Bessel functions $j_l$ look like damped sinusoids with an approximate phase difference of $\pi$ radians between $l$ and $l+1$, which leads to an antiphase relation between the HOMO and HOMO-1. As can be seen in Fig.~\ref{fig:Fig4}(b), this antiphase relation (between the HOMO-1 and HOMO or HOMO-2) is close to being realized, but not exact. In fact, following the work of Xu et al.~\cite{Xu1996a}, %given the experimental photon energy range and the C$_{60}$ radius,
a more quantitative study by Hasegawa et al.~\cite{Hasegawa1998a} introduced a modified asymptotic form for the Bessel functions with a correction factor $\alpha_{l_i}$ describing the initial and final states. In Hasegawa's work, the photoionization cross section is proportional to $\cos^2\big(kr+\alpha_{l_i}-\frac{l_i}{2}\pi\big)$ with $l_i=5$ and $\alpha_5 = 1.4$ for the HOMO and $l_i=4$ and $\alpha_4 = 1.0$ for the HOMO-1. We fit Hasegawa's model to our data (see Supp. Mat.) finding values of $\alpha_4=1.5$ and $\alpha_5=1.7$. More importantly, we find an approximate phase difference between the two band manifolds of $2.50$ rad which is very close to the phase difference found through our Fourier analysis of the raw data, $2.44$ rad 
%in the range $\{-\pi, \pi\}$ ($3.85$ rad in the range $\{0, 2\pi\}$)
as is illustrated in Fig.~\ref{fig:Fig4}(d).
%DL: Add in extracted phase values for the HOMO, HOMO-1 above.
%Discuss about HOMO and sigma / pi orbitals

The results from the FFT phase analysis of our data (Fig.~\ref{fig:Fig4}(c)) show that while the phase across the HOMO and HOMO-1 band manifolds is nearly constant with energy, it varies substantially across the energy range of the HOMO-2. This is shown in closer detail in Fig.~\ref{fig:Fig4}(d) where we plot the $k_y$-integrated (over the full window shown) phase for each band manifold. (For a similar plot see Fig. S5 Supp. Mat.). In order to more easily compare phase deviations from the mean within each band manifold, the zero is set to its mean phase for each band manifold (i.e. 3.03~rad, -0.82~rad, and 1.55~rad in the range $\{-\pi, \pi$\} for the HOMO, HOMO-1, and HOMO-2, respectively) as shown to the right of Fig. ~\ref{fig:Fig4}(d). The range of phase deviation within the HOMO-2 is clearly much larger than for the HOMO and HOMO-1. At $\Gamma$, the HOMO-2 phase (depending on energy) spans a range of 1.24~rad, while the equivalent values for the HOMO and HOMO-1 are only 0.14~rad and 0.22~rad, respectively. As shown to the right of panel (d), the top energies of the HOMO-2 (-4~eV to -4.43~eV) have a mean phase of 0.87~rad, the middle energies (-4.43~eV to -4.95~eV) have 1.40~rad, and the bottom energies (-4.95~eV to -5.8~eV) have 1.86~rad. The significant difference in phase across these energy ranges suggest a difference in orbital character of the states present within each of these energy ranges. This suggestion is strengthened by our previously discussed polarization-dependent observations for the middle and bottom energy ranges in HOMO-2. The middle energy hosts the weakly-dispersive bands with opposite dispersions using $\perp$ or $\parallel$ light polarization (Fig.~\ref{fig:Fig2}(c)), while the bottom energy has weaker bands that do not exhibit a clear linear-polarization-dependent effect. As for the circular dichroism studies, we see a clear flipping of the CD polarization between the middle (polarization of 0.10) and bottom (opposite pol. of 0.04) energy ranges (Fig.~\ref{fig:Fig3}(a)). These combined observations from our linear-polarization dependence, CD measurements, and photon-energy dependence strongly suggest a differing in orbital character of the bands within these energy ranges. The HOMO-2 is composed of $\pi_3$, $\sigma_8$, and $\sigma_9$ states \cite{Martins1991,Benning1991} (or possibly $\pi_3$, $\sigma_9$, and $\sigma_{10}$ \cite{Colavita2001}). In other words, the $\pi_3$, $\sigma_8$, and $\sigma_9$ bands are not spread evenly across the energy range, but instead are more strongly localized within certain energy ranges. Theory calculations support a localization in energy between the $\pi$ and $\sigma$ states, but are not definitive to their relative ordering in energy as certain models calculate the $\sigma$ states to lie at the top HOMO-2 energies \cite{Braga1991}, while others calculate the $\sigma$ states to lie near the bottom energies \cite{Hasegawa1998}, and others still show them nestled in the middle energies between the $\pi$ states \cite{Martins1991,Troullier1992}. 

Overall, the unusual phase deviations in the HOMO-2 are a result of the mixed orbital character of this band manifold. The continuous phase shift across the HOMO-2 might hide interesting phenomena that go beyond the discrete integer $l=5,4,3$ dependence (of the $\pi$ states) of the spherical Bessel functions that is associated to the simple interference models introduced before \cite{Hasegawa1998a,Xu1996a}. Wang et al.~\cite{Wang2008a} suggested that the HOMO-2 oscillations are almost entirely due to the $\pi$ states as they predict the $\sigma$ states will only affect the overall background due to deconstructive interference. Toffoli et al.~\cite{Toffoli2011} similarly predicted the $\sigma$ states to only contribute to a smooth background and not exhibit clear oscillations. The broader picture presented here, granted by our study of the linearly- and circularly-polarized-light dependence as well as photon-energy dependence, allow us to conclude that while the observations for the HOMO-2 are consistent with what is expected from the $\pi$ states, the $\sigma$ states also have a significant effect, in contrast to our HOMO and HOMO-1 observations (which have no $\sigma$ states). Further theoretical studies may provide further insight on the oscillations of the HOMO-2 band manifold intensity with photon energy.

One consequence of the observed strong effects of the particular photon energy and polarization on a molecular solid such as C$_{60}$ (where molecules have multiple degrees of freedom) is the proof of the absence of spinning of the molecules at low temperatures (20~K) and, most importantly, the constraint on the orientation of the C$_{60}$ molecules imposed by the substrate used (Bi$_2$Se$_3$), that as we have reported in the past \cite{Latzke2019} reduces the orientational disorder of a C$_{60}$ thin film. A similar study using varied photon energy and polarization can be applied to other molecular solids, providing a probe for the orbital character across individual valence band clusters. Additionally, by studying the effects of photon energy and polarization on the photoemission at different temperatures, new structural phase transitions characterized by a certain molecular orientation in a variety of molecular solids may be revealed. 

In conclusion, we have shown that incident light polarization and energy have strong, unique, and sometimes unexpected effects on C$_{60}$ photoemission which give us a precise view into the orbital character of the C$_{60}$ valence band structure. Concealed bands and band splittings are now visible offering a previously unseen view of the complex band structure within each band manifold. Previous and future studies must be made aware of the strong intensity effects that can affect the apparent dispersions observed and band locations extracted. We have expanded upon the photon-energy-dependent C$_{60}$ literature finding that a model based on k$_z$ rather than photon energy fits the oscillations very well while showing that the band orbital character is intimately coupled with the oscillation phase which has allowed for a more precise determination of the energy and momentum locations of the separate C$_{60}$ orbital characters (especially for the HOMO-2). These orbital character determinations provide us fundamental information about the relationship between molecular and crystalline electronic states in C$_{60}$.
%End Conclusion

\begin{acknowledgments}
The Advanced Light Source is supported by the Director, Office of Science, Office of Basic Energy Sciences, of the U.S.~Department of Energy (U.S.~DOE-BES) under contract No.~DE-AC02-05CH11231.   
The sp$^2$ Program (KC2207) provided the primary funding for this work.
C.O-A and R.R. were funded by the U.S.~DOE-BES under contract DE-SC0018154 for sample fabrication, ARPES data acquisition and data analysis. C.O-A would like to acknowledge fruitful discussions with Luca Moreschini
%Acknowledgments text.
%The ARPES work was supported by the sp2 Program at Lawrence Berkeley National Laboratory, funded by the U.S. Department of Energy, Office of Science, Office of Basic Energy Sciences, Materials Sciences and Engineering Division, under Contract No. DE-AC02-05CH11231. The Advanced Light Source is supported by the Director, Office of Science, Office of Basic Energy Sciences, of the U.S. Department of Energy under Contract No. DE-AC02-05CH11231.
\end{acknowledgments}

\bibliography{BiblioDWL}

\begin{thebibliography}{54}%
\makeatletter
\providecommand \@ifxundefined [1]{%
 \@ifx{#1\undefined}
}%
\providecommand \@ifnum [1]{%
 \ifnum #1\expandafter \@firstoftwo
 \else \expandafter \@secondoftwo
 \fi
}%
\providecommand \@ifx [1]{%
 \ifx #1\expandafter \@firstoftwo
 \else \expandafter \@secondoftwo
 \fi
}%
\providecommand \natexlab [1]{#1}%
\providecommand \enquote  [1]{``#1''}%
\providecommand \bibnamefont  [1]{#1}%
\providecommand \bibfnamefont [1]{#1}%
\providecommand \citenamefont [1]{#1}%
\providecommand \href@noop [0]{\@secondoftwo}%
\providecommand \href [0]{\begingroup \@sanitize@url \@href}%
\providecommand \@href[1]{\@@startlink{#1}\@@href}%
\providecommand \@@href[1]{\endgroup#1\@@endlink}%
\providecommand \@sanitize@url [0]{\catcode `\\12\catcode `\$12\catcode
  `\&12\catcode `\#12\catcode `\^12\catcode `\_12\catcode `\%12\relax}%
\providecommand \@@startlink[1]{}%
\providecommand \@@endlink[0]{}%
\providecommand \url  [0]{\begingroup\@sanitize@url \@url }%
\providecommand \@url [1]{\endgroup\@href {#1}{\urlprefix }}%
\providecommand \urlprefix  [0]{URL }%
\providecommand \Eprint [0]{\href }%
\providecommand \doibase [0]{http://dx.doi.org/}%
\providecommand \selectlanguage [0]{\@gobble}%
\providecommand \bibinfo  [0]{\@secondoftwo}%
\providecommand \bibfield  [0]{\@secondoftwo}%
\providecommand \translation [1]{[#1]}%
\providecommand \BibitemOpen [0]{}%
\providecommand \bibitemStop [0]{}%
\providecommand \bibitemNoStop [0]{.\EOS\space}%
\providecommand \EOS [0]{\spacefactor3000\relax}%
\providecommand \BibitemShut  [1]{\csname bibitem#1\endcsname}%
\let\auto@bib@innerbib\@empty
%</preamble>
\bibitem [{\citenamefont {Rao}\ \emph {et~al.}(2010)\citenamefont {Rao},
  \citenamefont {Wilson}, \citenamefont {Hodgkiss}, \citenamefont
  {Albert-Seifried}, \citenamefont {B{\"a}ssler},\ and\ \citenamefont
  {Friend}}]{Rao2010}%
  \BibitemOpen
  \bibfield  {author} {\bibinfo {author} {\bibfnamefont {A.}~\bibnamefont
  {Rao}}, \bibinfo {author} {\bibfnamefont {M.~W.~B.}\ \bibnamefont {Wilson}},
  \bibinfo {author} {\bibfnamefont {J.~M.}\ \bibnamefont {Hodgkiss}}, \bibinfo
  {author} {\bibfnamefont {S.}~\bibnamefont {Albert-Seifried}}, \bibinfo
  {author} {\bibfnamefont {H.}~\bibnamefont {B{\"a}ssler}}, \ and\ \bibinfo
  {author} {\bibfnamefont {R.~H.}\ \bibnamefont {Friend}},\ }\href {\doibase
  10.1021/ja1042462} {\bibfield  {journal} {\bibinfo  {journal} {Journal of the
  American Chemical Society}\ }\textbf {\bibinfo {volume} {132}},\ \bibinfo
  {pages} {12698} (\bibinfo {year} {2010})}\BibitemShut {NoStop}%
\bibitem [{\citenamefont {Liao}\ \emph {et~al.}(2014)\citenamefont {Liao},
  \citenamefont {Jhuo}, \citenamefont {Yeh}, \citenamefont {Cheng},
  \citenamefont {Li}, \citenamefont {Lee}, \citenamefont {Sharma},\ and\
  \citenamefont {Chen}}]{Liao2014}%
  \BibitemOpen
  \bibfield  {author} {\bibinfo {author} {\bibfnamefont {S.-H.}\ \bibnamefont
  {Liao}}, \bibinfo {author} {\bibfnamefont {H.-J.}\ \bibnamefont {Jhuo}},
  \bibinfo {author} {\bibfnamefont {P.-N.}\ \bibnamefont {Yeh}}, \bibinfo
  {author} {\bibfnamefont {Y.-S.}\ \bibnamefont {Cheng}}, \bibinfo {author}
  {\bibfnamefont {Y.-L.}\ \bibnamefont {Li}}, \bibinfo {author} {\bibfnamefont
  {Y.-H.}\ \bibnamefont {Lee}}, \bibinfo {author} {\bibfnamefont
  {S.}~\bibnamefont {Sharma}}, \ and\ \bibinfo {author} {\bibfnamefont {S.-A.}\
  \bibnamefont {Chen}},\ }\href {http://dx.doi.org/10.1038/srep06813}
  {\bibfield  {journal} {\bibinfo  {journal} {Scientific Reports}\ }\textbf
  {\bibinfo {volume} {4}},\ \bibinfo {pages} {6813} (\bibinfo {year}
  {2014})}\BibitemShut {NoStop}%
\bibitem [{\citenamefont {Liu}\ \emph {et~al.}(2014)\citenamefont {Liu},
  \citenamefont {Zhao}, \citenamefont {Li}, \citenamefont {Mu}, \citenamefont
  {Ma}, \citenamefont {Hu}, \citenamefont {Jiang}, \citenamefont {Lin},
  \citenamefont {Ade},\ and\ \citenamefont {Yan}}]{Liu2014}%
  \BibitemOpen
  \bibfield  {author} {\bibinfo {author} {\bibfnamefont {Y.}~\bibnamefont
  {Liu}}, \bibinfo {author} {\bibfnamefont {J.}~\bibnamefont {Zhao}}, \bibinfo
  {author} {\bibfnamefont {Z.}~\bibnamefont {Li}}, \bibinfo {author}
  {\bibfnamefont {C.}~\bibnamefont {Mu}}, \bibinfo {author} {\bibfnamefont
  {W.}~\bibnamefont {Ma}}, \bibinfo {author} {\bibfnamefont {H.}~\bibnamefont
  {Hu}}, \bibinfo {author} {\bibfnamefont {K.}~\bibnamefont {Jiang}}, \bibinfo
  {author} {\bibfnamefont {H.}~\bibnamefont {Lin}}, \bibinfo {author}
  {\bibfnamefont {H.}~\bibnamefont {Ade}}, \ and\ \bibinfo {author}
  {\bibfnamefont {H.}~\bibnamefont {Yan}},\ }\href {\doibase
  10.1038/ncomms6293} {\bibfield  {journal} {\bibinfo  {journal} {Nature
  Communications}\ }\textbf {\bibinfo {volume} {5}},\ \bibinfo {pages} {5293}
  (\bibinfo {year} {2014})}\BibitemShut {NoStop}%
\bibitem [{\citenamefont {Ojeda-Aristizabal}\ \emph {et~al.}(2017)\citenamefont
  {Ojeda-Aristizabal}, \citenamefont {Santos}, \citenamefont {Onishi},
  \citenamefont {Yan}, \citenamefont {Rasool}, \citenamefont {Kahn},
  \citenamefont {Lv}, \citenamefont {Latzke}, \citenamefont {Velasco},
  \citenamefont {Crommie}, \citenamefont {Sorensen}, \citenamefont {Gotlieb},
  \citenamefont {Lin}, \citenamefont {Watanabe}, \citenamefont {Taniguchi},
  \citenamefont {Lanzara},\ and\ \citenamefont
  {Zettl}}]{Ojeda-Aristizabal2017}%
  \BibitemOpen
  \bibfield  {author} {\bibinfo {author} {\bibfnamefont {C.}~\bibnamefont
  {Ojeda-Aristizabal}}, \bibinfo {author} {\bibfnamefont {E.~J.~G.}\
  \bibnamefont {Santos}}, \bibinfo {author} {\bibfnamefont {S.}~\bibnamefont
  {Onishi}}, \bibinfo {author} {\bibfnamefont {A.}~\bibnamefont {Yan}},
  \bibinfo {author} {\bibfnamefont {H.~I.}\ \bibnamefont {Rasool}}, \bibinfo
  {author} {\bibfnamefont {S.}~\bibnamefont {Kahn}}, \bibinfo {author}
  {\bibfnamefont {Y.}~\bibnamefont {Lv}}, \bibinfo {author} {\bibfnamefont
  {D.~W.}\ \bibnamefont {Latzke}}, \bibinfo {author} {\bibfnamefont
  {J.}~\bibnamefont {Velasco}}, \bibinfo {author} {\bibfnamefont {M.~F.}\
  \bibnamefont {Crommie}}, \bibinfo {author} {\bibfnamefont {M.}~\bibnamefont
  {Sorensen}}, \bibinfo {author} {\bibfnamefont {K.}~\bibnamefont {Gotlieb}},
  \bibinfo {author} {\bibfnamefont {C.-Y.}\ \bibnamefont {Lin}}, \bibinfo
  {author} {\bibfnamefont {K.}~\bibnamefont {Watanabe}}, \bibinfo {author}
  {\bibfnamefont {T.}~\bibnamefont {Taniguchi}}, \bibinfo {author}
  {\bibfnamefont {A.}~\bibnamefont {Lanzara}}, \ and\ \bibinfo {author}
  {\bibfnamefont {A.}~\bibnamefont {Zettl}},\ }\bibfield  {booktitle} {\emph
  {\bibinfo {booktitle} {ACS Nano}},\ }\href {\doibase 10.1021/acsnano.7b00551}
  {\bibfield  {journal} {\bibinfo  {journal} {ACS Nano}\ }\textbf {\bibinfo
  {volume} {11}},\ \bibinfo {pages} {4686} (\bibinfo {year}
  {2017})}\BibitemShut {NoStop}%
\bibitem [{\citenamefont {Kim}\ \emph {et~al.}(2015)\citenamefont {Kim},
  \citenamefont {Lee}, \citenamefont {Santos}, \citenamefont {Jo},
  \citenamefont {Salleo}, \citenamefont {Nishi},\ and\ \citenamefont
  {Bao}}]{Kim2015}%
  \BibitemOpen
  \bibfield  {author} {\bibinfo {author} {\bibfnamefont {K.}~\bibnamefont
  {Kim}}, \bibinfo {author} {\bibfnamefont {T.~H.}\ \bibnamefont {Lee}},
  \bibinfo {author} {\bibfnamefont {E.~J.~G.}\ \bibnamefont {Santos}}, \bibinfo
  {author} {\bibfnamefont {P.~S.}\ \bibnamefont {Jo}}, \bibinfo {author}
  {\bibfnamefont {A.}~\bibnamefont {Salleo}}, \bibinfo {author} {\bibfnamefont
  {Y.}~\bibnamefont {Nishi}}, \ and\ \bibinfo {author} {\bibfnamefont
  {Z.}~\bibnamefont {Bao}},\ }\href {\doibase 10.1021/acsnano.5b00581}
  {\bibfield  {journal} {\bibinfo  {journal} {{ACS} Nano}\ }\textbf {\bibinfo
  {volume} {9}},\ \bibinfo {pages} {5922} (\bibinfo {year} {2015})}\BibitemShut
  {NoStop}%
\bibitem [{\citenamefont {Cami}\ \emph {et~al.}(2010)\citenamefont {Cami},
  \citenamefont {Bernard-Salas}, \citenamefont {Peeters},\ and\ \citenamefont
  {Malek}}]{Cami2010}%
  \BibitemOpen
  \bibfield  {author} {\bibinfo {author} {\bibfnamefont {J.}~\bibnamefont
  {Cami}}, \bibinfo {author} {\bibfnamefont {J.}~\bibnamefont {Bernard-Salas}},
  \bibinfo {author} {\bibfnamefont {E.}~\bibnamefont {Peeters}}, \ and\
  \bibinfo {author} {\bibfnamefont {S.~E.}\ \bibnamefont {Malek}},\ }\href
  {\doibase 10.1126/science.1192035} {\bibfield  {journal} {\bibinfo  {journal}
  {Science}\ }\textbf {\bibinfo {volume} {329}},\ \bibinfo {pages} {1180}
  (\bibinfo {year} {2010})}\BibitemShut {NoStop}%
\bibitem [{\citenamefont {Sellgren}\ \emph {et~al.}(2010)\citenamefont
  {Sellgren}, \citenamefont {Werner}, \citenamefont {Ingalls}, \citenamefont
  {Smith}, \citenamefont {Carleton},\ and\ \citenamefont
  {Joblin}}]{Sellgren2010}%
  \BibitemOpen
  \bibfield  {author} {\bibinfo {author} {\bibfnamefont {K.}~\bibnamefont
  {Sellgren}}, \bibinfo {author} {\bibfnamefont {M.~W.}\ \bibnamefont
  {Werner}}, \bibinfo {author} {\bibfnamefont {J.~G.}\ \bibnamefont {Ingalls}},
  \bibinfo {author} {\bibfnamefont {J.~D.~T.}\ \bibnamefont {Smith}}, \bibinfo
  {author} {\bibfnamefont {T.~M.}\ \bibnamefont {Carleton}}, \ and\ \bibinfo
  {author} {\bibfnamefont {C.}~\bibnamefont {Joblin}},\ }\href {\doibase
  10.1088/2041-8205/722/1/l54} {\bibfield  {journal} {\bibinfo  {journal} {The
  Astrophysical Journal}\ }\textbf {\bibinfo {volume} {722}},\ \bibinfo {pages}
  {L54} (\bibinfo {year} {2010})}\BibitemShut {NoStop}%
\bibitem [{\citenamefont {Campbell}\ \emph {et~al.}(2015)\citenamefont
  {Campbell}, \citenamefont {Holz}, \citenamefont {Gerlich},\ and\
  \citenamefont {Maier}}]{Campbell2015}%
  \BibitemOpen
  \bibfield  {author} {\bibinfo {author} {\bibfnamefont {E.~K.}\ \bibnamefont
  {Campbell}}, \bibinfo {author} {\bibfnamefont {M.}~\bibnamefont {Holz}},
  \bibinfo {author} {\bibfnamefont {D.}~\bibnamefont {Gerlich}}, \ and\
  \bibinfo {author} {\bibfnamefont {J.~P.}\ \bibnamefont {Maier}},\ }\href
  {http://dx.doi.org/10.1038/nature14566} {\bibfield  {journal} {\bibinfo
  {journal} {Nature}\ }\textbf {\bibinfo {volume} {523}},\ \bibinfo {pages}
  {322} (\bibinfo {year} {2015})}\BibitemShut {NoStop}%
\bibitem [{\citenamefont {Latzke}\ \emph {et~al.}(2019)\citenamefont {Latzke},
  \citenamefont {Ojeda-Aristizabal}, \citenamefont {Griffin}, \citenamefont
  {Denlinger}, \citenamefont {Neaton}, \citenamefont {Zettl},\ and\
  \citenamefont {Lanzara}}]{Latzke2019}%
  \BibitemOpen
  \bibfield  {author} {\bibinfo {author} {\bibfnamefont {D.~W.}\ \bibnamefont
  {Latzke}}, \bibinfo {author} {\bibfnamefont {C.}~\bibnamefont
  {Ojeda-Aristizabal}}, \bibinfo {author} {\bibfnamefont {S.~M.}\ \bibnamefont
  {Griffin}}, \bibinfo {author} {\bibfnamefont {J.~D.}\ \bibnamefont
  {Denlinger}}, \bibinfo {author} {\bibfnamefont {J.~B.}\ \bibnamefont
  {Neaton}}, \bibinfo {author} {\bibfnamefont {A.}~\bibnamefont {Zettl}}, \
  and\ \bibinfo {author} {\bibfnamefont {A.}~\bibnamefont {Lanzara}},\ }\href
  {\doibase 10.1103/physrevb.99.045425} {\bibfield  {journal} {\bibinfo
  {journal} {Physical Review B}\ }\textbf {\bibinfo {volume} {99}},\ \bibinfo
  {pages} {045425} (\bibinfo {year} {2019})}\BibitemShut {NoStop}%
\bibitem [{\citenamefont {Dresselhaus}\ \emph {et~al.}(1996)\citenamefont
  {Dresselhaus}, \citenamefont {Dresselhaus},\ and\ \citenamefont
  {Eklund}}]{Dresselhaus1996}%
  \BibitemOpen
  \bibfield  {author} {\bibinfo {author} {\bibfnamefont {M.}~\bibnamefont
  {Dresselhaus}}, \bibinfo {author} {\bibfnamefont {G.}~\bibnamefont
  {Dresselhaus}}, \ and\ \bibinfo {author} {\bibfnamefont {P.}~\bibnamefont
  {Eklund}},\ }\href {https://books.google.com/books?id=T8NLqyOMZ50C} {\emph
  {\bibinfo {title} {Science of Fullerenes and Carbon Nanotubes: Their
  Properties and Applications}}}\ (\bibinfo  {publisher} {Elsevier Science},\
  \bibinfo {year} {1996})\BibitemShut {NoStop}%
\bibitem [{\citenamefont {Korm\'anyos}\ \emph {et~al.}(2013)\citenamefont
  {Korm\'anyos}, \citenamefont {Z\'olyomi}, \citenamefont {Drummond},
  \citenamefont {Rakyta}, \citenamefont {Burkard},\ and\ \citenamefont
  {Fal'ko}}]{Kormanyos2013}%
  \BibitemOpen
  \bibfield  {author} {\bibinfo {author} {\bibfnamefont {A.}~\bibnamefont
  {Korm\'anyos}}, \bibinfo {author} {\bibfnamefont {V.}~\bibnamefont
  {Z\'olyomi}}, \bibinfo {author} {\bibfnamefont {N.~D.}\ \bibnamefont
  {Drummond}}, \bibinfo {author} {\bibfnamefont {P.}~\bibnamefont {Rakyta}},
  \bibinfo {author} {\bibfnamefont {G.}~\bibnamefont {Burkard}}, \ and\
  \bibinfo {author} {\bibfnamefont {V.~I.}\ \bibnamefont {Fal'ko}},\ }\href
  {\doibase 10.1103/PhysRevB.88.045416} {\bibfield  {journal} {\bibinfo
  {journal} {Phys. Rev. B}\ }\textbf {\bibinfo {volume} {88}},\ \bibinfo
  {pages} {045416} (\bibinfo {year} {2013})}\BibitemShut {NoStop}%
\bibitem [{\citenamefont {Rostami}\ \emph {et~al.}(2013)\citenamefont
  {Rostami}, \citenamefont {Moghaddam},\ and\ \citenamefont
  {Asgari}}]{Rostami2013}%
  \BibitemOpen
  \bibfield  {author} {\bibinfo {author} {\bibfnamefont {H.}~\bibnamefont
  {Rostami}}, \bibinfo {author} {\bibfnamefont {A.~G.}\ \bibnamefont
  {Moghaddam}}, \ and\ \bibinfo {author} {\bibfnamefont {R.}~\bibnamefont
  {Asgari}},\ }\href {http://link.aps.org/doi/10.1103/PhysRevB.88.085440}
  {\bibfield  {journal} {\bibinfo  {journal} {Phys. Rev. B}\ }\textbf {\bibinfo
  {volume} {88}},\ \bibinfo {pages} {085440} (\bibinfo {year}
  {2013})}\BibitemShut {NoStop}%
\bibitem [{\citenamefont {Zhou}\ \emph {et~al.}(2006)\citenamefont {Zhou},
  \citenamefont {Gweon},\ and\ \citenamefont {Lanzara}}]{Zhou2006a}%
  \BibitemOpen
  \bibfield  {author} {\bibinfo {author} {\bibfnamefont {S.~Y.}\ \bibnamefont
  {Zhou}}, \bibinfo {author} {\bibfnamefont {G.-H.}\ \bibnamefont {Gweon}}, \
  and\ \bibinfo {author} {\bibfnamefont {A.}~\bibnamefont {Lanzara}},\ }\href
  {\doibase 10.1016/j.aop.2006.04.011} {\bibfield  {journal} {\bibinfo
  {journal} {Annals of Physics}\ }\textbf {\bibinfo {volume} {321}},\ \bibinfo
  {pages} {1730} (\bibinfo {year} {2006})}\BibitemShut {NoStop}%
\bibitem [{\citenamefont {Thirupathaiah}\ \emph {et~al.}(2010)\citenamefont
  {Thirupathaiah}, \citenamefont {de~Jong}, \citenamefont {Ovsyannikov},
  \citenamefont {Dürr}, \citenamefont {Varykhalov}, \citenamefont {Follath},
  \citenamefont {Huang}, \citenamefont {Huisman}, \citenamefont {Golden},
  \citenamefont {Zhang}, \citenamefont {Jeschke}, \citenamefont
  {Valent{\'{\i}}}, \citenamefont {Erb}, \citenamefont {Gloskovskii},\ and\
  \citenamefont {Fink}}]{Thirupathaiah2010}%
  \BibitemOpen
  \bibfield  {author} {\bibinfo {author} {\bibfnamefont {S.}~\bibnamefont
  {Thirupathaiah}}, \bibinfo {author} {\bibfnamefont {S.}~\bibnamefont
  {de~Jong}}, \bibinfo {author} {\bibfnamefont {R.}~\bibnamefont
  {Ovsyannikov}}, \bibinfo {author} {\bibfnamefont {H.~A.}\ \bibnamefont
  {Dürr}}, \bibinfo {author} {\bibfnamefont {A.}~\bibnamefont {Varykhalov}},
  \bibinfo {author} {\bibfnamefont {R.}~\bibnamefont {Follath}}, \bibinfo
  {author} {\bibfnamefont {Y.}~\bibnamefont {Huang}}, \bibinfo {author}
  {\bibfnamefont {R.}~\bibnamefont {Huisman}}, \bibinfo {author} {\bibfnamefont
  {M.~S.}\ \bibnamefont {Golden}}, \bibinfo {author} {\bibfnamefont {Y.-Z.}\
  \bibnamefont {Zhang}}, \bibinfo {author} {\bibfnamefont {H.~O.}\ \bibnamefont
  {Jeschke}}, \bibinfo {author} {\bibfnamefont {R.}~\bibnamefont
  {Valent{\'{\i}}}}, \bibinfo {author} {\bibfnamefont {A.}~\bibnamefont {Erb}},
  \bibinfo {author} {\bibfnamefont {A.}~\bibnamefont {Gloskovskii}}, \ and\
  \bibinfo {author} {\bibfnamefont {J.}~\bibnamefont {Fink}},\ }\href {\doibase
  10.1103/physrevb.81.104512} {\bibfield  {journal} {\bibinfo  {journal}
  {Physical Review B}\ }\textbf {\bibinfo {volume} {81}},\ \bibinfo {pages}
  {104512} (\bibinfo {year} {2010})}\BibitemShut {NoStop}%
\bibitem [{\citenamefont {Wang}\ \emph {et~al.}(2012)\citenamefont {Wang},
  \citenamefont {Richard}, \citenamefont {Huang}, \citenamefont {Miao},
  \citenamefont {Cevey}, \citenamefont {Xu}, \citenamefont {Sun}, \citenamefont
  {Qian}, \citenamefont {Xu}, \citenamefont {Shi}, \citenamefont {Hu},
  \citenamefont {Dai},\ and\ \citenamefont {Ding}}]{Wang2012e}%
  \BibitemOpen
  \bibfield  {author} {\bibinfo {author} {\bibfnamefont {X.-P.}\ \bibnamefont
  {Wang}}, \bibinfo {author} {\bibfnamefont {P.}~\bibnamefont {Richard}},
  \bibinfo {author} {\bibfnamefont {Y.-B.}\ \bibnamefont {Huang}}, \bibinfo
  {author} {\bibfnamefont {H.}~\bibnamefont {Miao}}, \bibinfo {author}
  {\bibfnamefont {L.}~\bibnamefont {Cevey}}, \bibinfo {author} {\bibfnamefont
  {N.}~\bibnamefont {Xu}}, \bibinfo {author} {\bibfnamefont {Y.-J.}\
  \bibnamefont {Sun}}, \bibinfo {author} {\bibfnamefont {T.}~\bibnamefont
  {Qian}}, \bibinfo {author} {\bibfnamefont {Y.-M.}\ \bibnamefont {Xu}},
  \bibinfo {author} {\bibfnamefont {M.}~\bibnamefont {Shi}}, \bibinfo {author}
  {\bibfnamefont {J.-P.}\ \bibnamefont {Hu}}, \bibinfo {author} {\bibfnamefont
  {X.}~\bibnamefont {Dai}}, \ and\ \bibinfo {author} {\bibfnamefont
  {H.}~\bibnamefont {Ding}},\ }\href {\doibase 10.1103/physrevb.85.214518}
  {\bibfield  {journal} {\bibinfo  {journal} {Physical Review B}\ }\textbf
  {\bibinfo {volume} {85}},\ \bibinfo {pages} {214518} (\bibinfo {year}
  {2012})}\BibitemShut {NoStop}%
\bibitem [{\citenamefont {Lichtenberger}\ \emph {et~al.}(1993)\citenamefont
  {Lichtenberger}, \citenamefont {Wright}, \citenamefont {Gruhn},\ and\
  \citenamefont {Rempe}}]{Lichtenberger1993}%
  \BibitemOpen
  \bibfield  {author} {\bibinfo {author} {\bibfnamefont {D.~L.}\ \bibnamefont
  {Lichtenberger}}, \bibinfo {author} {\bibfnamefont {L.~L.}\ \bibnamefont
  {Wright}}, \bibinfo {author} {\bibfnamefont {N.~E.}\ \bibnamefont {Gruhn}}, \
  and\ \bibinfo {author} {\bibfnamefont {M.~E.}\ \bibnamefont {Rempe}},\ }\href
  {\doibase 10.1016/0379-6779(93)91167-z} {\bibfield  {journal} {\bibinfo
  {journal} {Synthetic Metals}\ }\textbf {\bibinfo {volume} {59}},\ \bibinfo
  {pages} {353} (\bibinfo {year} {1993})}\BibitemShut {NoStop}%
\bibitem [{\citenamefont {Vos}\ \emph {et~al.}(1997)\citenamefont {Vos},
  \citenamefont {Canney}, \citenamefont {McCarthy}, \citenamefont {Utteridge},
  \citenamefont {Michalewicz},\ and\ \citenamefont {Weigold}}]{Vos1997}%
  \BibitemOpen
  \bibfield  {author} {\bibinfo {author} {\bibfnamefont {M.}~\bibnamefont
  {Vos}}, \bibinfo {author} {\bibfnamefont {S.~A.}\ \bibnamefont {Canney}},
  \bibinfo {author} {\bibfnamefont {I.~E.}\ \bibnamefont {McCarthy}}, \bibinfo
  {author} {\bibfnamefont {S.}~\bibnamefont {Utteridge}}, \bibinfo {author}
  {\bibfnamefont {M.~T.}\ \bibnamefont {Michalewicz}}, \ and\ \bibinfo {author}
  {\bibfnamefont {E.}~\bibnamefont {Weigold}},\ }\href {\doibase
  10.1103/physrevb.56.1309} {\bibfield  {journal} {\bibinfo  {journal}
  {Physical Review B}\ }\textbf {\bibinfo {volume} {56}},\ \bibinfo {pages}
  {1309} (\bibinfo {year} {1997})}\BibitemShut {NoStop}%
\bibitem [{\citenamefont {Schiessling}\ \emph {et~al.}(2003)\citenamefont
  {Schiessling}, \citenamefont {Kjeldgaard}, \citenamefont {Balasubramanian},
  \citenamefont {Nordgren},\ and\ \citenamefont
  {Br\"uhwiler}}]{Schiessling2003}%
  \BibitemOpen
  \bibfield  {author} {\bibinfo {author} {\bibfnamefont {J.}~\bibnamefont
  {Schiessling}}, \bibinfo {author} {\bibfnamefont {L.}~\bibnamefont
  {Kjeldgaard}}, \bibinfo {author} {\bibfnamefont {T.}~\bibnamefont
  {Balasubramanian}}, \bibinfo {author} {\bibfnamefont {J.}~\bibnamefont
  {Nordgren}}, \ and\ \bibinfo {author} {\bibfnamefont {P.~A.}\ \bibnamefont
  {Br\"uhwiler}},\ }\href {\doibase 10.1103/PhysRevB.68.205405} {\bibfield
  {journal} {\bibinfo  {journal} {Phys. Rev. B}\ }\textbf {\bibinfo {volume}
  {68}},\ \bibinfo {pages} {205405} (\bibinfo {year} {2003})}\BibitemShut
  {NoStop}%
\bibitem [{\citenamefont {Fecher}\ \emph {et~al.}(1997)\citenamefont {Fecher},
  \citenamefont {Gr{\"u}newald}, \citenamefont {Merkel}, \citenamefont
  {Ostertag}, \citenamefont {Oelsner}, \citenamefont {Sch{\"o}nhense},
  \citenamefont {Jentzsch},\ and\ \citenamefont {J{\"u}pner}}]{Fecher1997}%
  \BibitemOpen
  \bibfield  {author} {\bibinfo {author} {\bibfnamefont {G.}~\bibnamefont
  {Fecher}}, \bibinfo {author} {\bibfnamefont {C.}~\bibnamefont
  {Gr{\"u}newald}}, \bibinfo {author} {\bibfnamefont {M.}~\bibnamefont
  {Merkel}}, \bibinfo {author} {\bibfnamefont {C.}~\bibnamefont {Ostertag}},
  \bibinfo {author} {\bibfnamefont {A.}~\bibnamefont {Oelsner}}, \bibinfo
  {author} {\bibfnamefont {G.}~\bibnamefont {Sch{\"o}nhense}}, \bibinfo
  {author} {\bibfnamefont {T.}~\bibnamefont {Jentzsch}}, \ and\ \bibinfo
  {author} {\bibfnamefont {H.}~\bibnamefont {J{\"u}pner}},\ }\href {\doibase
  10.1016/s0040-6090(97)00088-6} {\bibfield  {journal} {\bibinfo  {journal}
  {Thin Solid Films}\ }\textbf {\bibinfo {volume} {303}},\ \bibinfo {pages}
  {58} (\bibinfo {year} {1997})}\BibitemShut {NoStop}%
\bibitem [{\citenamefont {R{\"u}del}\ \emph {et~al.}(2002)\citenamefont
  {R{\"u}del}, \citenamefont {Hentges}, \citenamefont {Becker}, \citenamefont
  {Chakraborty}, \citenamefont {Madjet},\ and\ \citenamefont
  {Rost}}]{Ruedel2002}%
  \BibitemOpen
  \bibfield  {author} {\bibinfo {author} {\bibfnamefont {A.}~\bibnamefont
  {R{\"u}del}}, \bibinfo {author} {\bibfnamefont {R.}~\bibnamefont {Hentges}},
  \bibinfo {author} {\bibfnamefont {U.}~\bibnamefont {Becker}}, \bibinfo
  {author} {\bibfnamefont {H.~S.}\ \bibnamefont {Chakraborty}}, \bibinfo
  {author} {\bibfnamefont {M.~E.}\ \bibnamefont {Madjet}}, \ and\ \bibinfo
  {author} {\bibfnamefont {J.~M.}\ \bibnamefont {Rost}},\ }\href {\doibase
  10.1103/physrevlett.89.125503} {\bibfield  {journal} {\bibinfo  {journal}
  {Physical Review Letters}\ }\textbf {\bibinfo {volume} {89}} (\bibinfo {year}
  {2002}),\ 10.1103/physrevlett.89.125503}\BibitemShut {NoStop}%
\bibitem [{\citenamefont {Becker}\ \emph {et~al.}(2000)\citenamefont {Becker},
  \citenamefont {Gessner},\ and\ \citenamefont {R{\"u}del}}]{Becker2000}%
  \BibitemOpen
  \bibfield  {author} {\bibinfo {author} {\bibfnamefont {U.}~\bibnamefont
  {Becker}}, \bibinfo {author} {\bibfnamefont {O.}~\bibnamefont {Gessner}}, \
  and\ \bibinfo {author} {\bibfnamefont {A.}~\bibnamefont {R{\"u}del}},\ }\href
  {\doibase 10.1016/s0368-2048(00)00128-6} {\bibfield  {journal} {\bibinfo
  {journal} {Journal of Electron Spectroscopy and Related Phenomena}\ }\textbf
  {\bibinfo {volume} {108}},\ \bibinfo {pages} {189} (\bibinfo {year}
  {2000})}\BibitemShut {NoStop}%
\bibitem [{\citenamefont {Frank}\ and\ \citenamefont {Rost}(1997)}]{Frank1997}%
  \BibitemOpen
  \bibfield  {author} {\bibinfo {author} {\bibfnamefont {O.}~\bibnamefont
  {Frank}}\ and\ \bibinfo {author} {\bibfnamefont {J.-M.}\ \bibnamefont
  {Rost}},\ }\href {\doibase 10.1016/s0009-2614(97)00471-5} {\bibfield
  {journal} {\bibinfo  {journal} {Chemical Physics Letters}\ }\textbf {\bibinfo
  {volume} {271}},\ \bibinfo {pages} {367} (\bibinfo {year}
  {1997})}\BibitemShut {NoStop}%
\bibitem [{\citenamefont {Xu}\ \emph {et~al.}(1996)\citenamefont {Xu},
  \citenamefont {Tan},\ and\ \citenamefont {Becker}}]{Xu1996a}%
  \BibitemOpen
  \bibfield  {author} {\bibinfo {author} {\bibfnamefont {Y.~B.}\ \bibnamefont
  {Xu}}, \bibinfo {author} {\bibfnamefont {M.~Q.}\ \bibnamefont {Tan}}, \ and\
  \bibinfo {author} {\bibfnamefont {U.}~\bibnamefont {Becker}},\ }\href
  {\doibase 10.1103/physrevlett.76.3538} {\bibfield  {journal} {\bibinfo
  {journal} {Physical Review Letters}\ }\textbf {\bibinfo {volume} {76}},\
  \bibinfo {pages} {3538} (\bibinfo {year} {1996})}\BibitemShut {NoStop}%
\bibitem [{\citenamefont {Benning}\ \emph {et~al.}(1991)\citenamefont
  {Benning}, \citenamefont {Poirier}, \citenamefont {Troullier}, \citenamefont
  {Martins}, \citenamefont {Weaver}, \citenamefont {Haufler}, \citenamefont
  {Chibante},\ and\ \citenamefont {Smalley}}]{Benning1991}%
  \BibitemOpen
  \bibfield  {author} {\bibinfo {author} {\bibfnamefont {P.~J.}\ \bibnamefont
  {Benning}}, \bibinfo {author} {\bibfnamefont {D.~M.}\ \bibnamefont
  {Poirier}}, \bibinfo {author} {\bibfnamefont {N.}~\bibnamefont {Troullier}},
  \bibinfo {author} {\bibfnamefont {J.~L.}\ \bibnamefont {Martins}}, \bibinfo
  {author} {\bibfnamefont {J.~H.}\ \bibnamefont {Weaver}}, \bibinfo {author}
  {\bibfnamefont {R.~E.}\ \bibnamefont {Haufler}}, \bibinfo {author}
  {\bibfnamefont {L.~P.~F.}\ \bibnamefont {Chibante}}, \ and\ \bibinfo {author}
  {\bibfnamefont {R.~E.}\ \bibnamefont {Smalley}},\ }\href {\doibase
  10.1103/physrevb.44.1962} {\bibfield  {journal} {\bibinfo  {journal}
  {Physical Review B}\ }\textbf {\bibinfo {volume} {44}},\ \bibinfo {pages}
  {1962} (\bibinfo {year} {1991})}\BibitemShut {NoStop}%
\bibitem [{\citenamefont {Wu}\ \emph {et~al.}(1992)\citenamefont {Wu},
  \citenamefont {Shen}, \citenamefont {Dessau}, \citenamefont {Cao},
  \citenamefont {Marshall}, \citenamefont {Pianetta}, \citenamefont {Lindau},
  \citenamefont {Yang}, \citenamefont {Terry}, \citenamefont {King},
  \citenamefont {Wells}, \citenamefont {Elloway}, \citenamefont {Wendt},
  \citenamefont {Brown}, \citenamefont {Hunziker},\ and\ \citenamefont
  {de~Vries}}]{Wu1992}%
  \BibitemOpen
  \bibfield  {author} {\bibinfo {author} {\bibfnamefont {J.}~\bibnamefont
  {Wu}}, \bibinfo {author} {\bibfnamefont {Z.-X.}\ \bibnamefont {Shen}},
  \bibinfo {author} {\bibfnamefont {D.}~\bibnamefont {Dessau}}, \bibinfo
  {author} {\bibfnamefont {R.}~\bibnamefont {Cao}}, \bibinfo {author}
  {\bibfnamefont {D.}~\bibnamefont {Marshall}}, \bibinfo {author}
  {\bibfnamefont {P.}~\bibnamefont {Pianetta}}, \bibinfo {author}
  {\bibfnamefont {I.}~\bibnamefont {Lindau}}, \bibinfo {author} {\bibfnamefont
  {X.}~\bibnamefont {Yang}}, \bibinfo {author} {\bibfnamefont {J.}~\bibnamefont
  {Terry}}, \bibinfo {author} {\bibfnamefont {D.}~\bibnamefont {King}},
  \bibinfo {author} {\bibfnamefont {B.}~\bibnamefont {Wells}}, \bibinfo
  {author} {\bibfnamefont {D.}~\bibnamefont {Elloway}}, \bibinfo {author}
  {\bibfnamefont {H.}~\bibnamefont {Wendt}}, \bibinfo {author} {\bibfnamefont
  {C.}~\bibnamefont {Brown}}, \bibinfo {author} {\bibfnamefont
  {H.}~\bibnamefont {Hunziker}}, \ and\ \bibinfo {author} {\bibfnamefont
  {M.}~\bibnamefont {de~Vries}},\ }\href
  {http://www.sciencedirect.com/science/article/pii/092145349290007Y}
  {\bibfield  {journal} {\bibinfo  {journal} {Physica C: Superconductivity}\
  }\textbf {\bibinfo {volume} {197}},\ \bibinfo {pages} {251} (\bibinfo {year}
  {1992})}\BibitemShut {NoStop}%
\bibitem [{\citenamefont {Wang}\ \emph {et~al.}(2008)\citenamefont {Wang},
  \citenamefont {Xu}, \citenamefont {Li}, \citenamefont {Zhang},\ and\
  \citenamefont {Xu}}]{Wang2008a}%
  \BibitemOpen
  \bibfield  {author} {\bibinfo {author} {\bibfnamefont {X.-X.}\ \bibnamefont
  {Wang}}, \bibinfo {author} {\bibfnamefont {Y.-B.}\ \bibnamefont {Xu}},
  \bibinfo {author} {\bibfnamefont {H.-N.}\ \bibnamefont {Li}}, \bibinfo
  {author} {\bibfnamefont {W.-H.}\ \bibnamefont {Zhang}}, \ and\ \bibinfo
  {author} {\bibfnamefont {F.-Q.}\ \bibnamefont {Xu}},\ }\href {\doibase
  10.1016/j.elspec.2008.07.004} {\bibfield  {journal} {\bibinfo  {journal}
  {Journal of Electron Spectroscopy and Related Phenomena}\ }\textbf {\bibinfo
  {volume} {165}},\ \bibinfo {pages} {20} (\bibinfo {year} {2008})}\BibitemShut
  {NoStop}%
\bibitem [{\citenamefont {Toffoli}\ \emph {et~al.}(2011)\citenamefont
  {Toffoli}, \citenamefont {Stener}, \citenamefont {Fronzoni},\ and\
  \citenamefont {Decleva}}]{Toffoli2011}%
  \BibitemOpen
  \bibfield  {author} {\bibinfo {author} {\bibfnamefont {D.}~\bibnamefont
  {Toffoli}}, \bibinfo {author} {\bibfnamefont {M.}~\bibnamefont {Stener}},
  \bibinfo {author} {\bibfnamefont {G.}~\bibnamefont {Fronzoni}}, \ and\
  \bibinfo {author} {\bibfnamefont {P.}~\bibnamefont {Decleva}},\ }\href
  {\doibase 10.1016/j.cplett.2011.10.006} {\bibfield  {journal} {\bibinfo
  {journal} {Chemical Physics Letters}\ }\textbf {\bibinfo {volume} {516}},\
  \bibinfo {pages} {154} (\bibinfo {year} {2011})}\BibitemShut {NoStop}%
\bibitem [{\citenamefont {Colavita}\ \emph {et~al.}(2001)\citenamefont
  {Colavita}, \citenamefont {De~Alti}, \citenamefont {Fronzoni}, \citenamefont
  {Stener},\ and\ \citenamefont {Decleva}}]{Colavita2001}%
  \BibitemOpen
  \bibfield  {author} {\bibinfo {author} {\bibfnamefont {P.}~\bibnamefont
  {Colavita}}, \bibinfo {author} {\bibfnamefont {G.}~\bibnamefont {De~Alti}},
  \bibinfo {author} {\bibfnamefont {G.}~\bibnamefont {Fronzoni}}, \bibinfo
  {author} {\bibfnamefont {M.}~\bibnamefont {Stener}}, \ and\ \bibinfo {author}
  {\bibfnamefont {P.}~\bibnamefont {Decleva}},\ }\href {\doibase
  10.1039/B104761M} {\bibfield  {journal} {\bibinfo  {journal} {Phys. Chem.
  Chem. Phys.}\ }\textbf {\bibinfo {volume} {3}},\ \bibinfo {pages} {4481}
  (\bibinfo {year} {2001})}\BibitemShut {NoStop}%
\bibitem [{\citenamefont {Hasegawa}\ \emph
  {et~al.}(1998{\natexlab{a}})\citenamefont {Hasegawa}, \citenamefont
  {Miyamae}, \citenamefont {Yakushi}, \citenamefont {Inokuchi}, \citenamefont
  {Seki},\ and\ \citenamefont {Ueno}}]{Hasegawa1998a}%
  \BibitemOpen
  \bibfield  {author} {\bibinfo {author} {\bibfnamefont {S.}~\bibnamefont
  {Hasegawa}}, \bibinfo {author} {\bibfnamefont {T.}~\bibnamefont {Miyamae}},
  \bibinfo {author} {\bibfnamefont {K.}~\bibnamefont {Yakushi}}, \bibinfo
  {author} {\bibfnamefont {H.}~\bibnamefont {Inokuchi}}, \bibinfo {author}
  {\bibfnamefont {K.}~\bibnamefont {Seki}}, \ and\ \bibinfo {author}
  {\bibfnamefont {N.}~\bibnamefont {Ueno}},\ }\href {\doibase
  10.1103/PhysRevB.58.4927} {\bibfield  {journal} {\bibinfo  {journal} {Phys.
  Rev. B}\ }\textbf {\bibinfo {volume} {58}},\ \bibinfo {pages} {4927}
  (\bibinfo {year} {1998}{\natexlab{a}})}\BibitemShut {NoStop}%
\bibitem [{\citenamefont {Liebsch}\ \emph {et~al.}(1995)\citenamefont
  {Liebsch}, \citenamefont {Plotzke}, \citenamefont {Heiser}, \citenamefont
  {Hergenhahn}, \citenamefont {Hemmers}, \citenamefont {Wehlitz}, \citenamefont
  {Viefhaus}, \citenamefont {Langer}, \citenamefont {Whitfield},\ and\
  \citenamefont {Becker}}]{Liebsch1995}%
  \BibitemOpen
  \bibfield  {author} {\bibinfo {author} {\bibfnamefont {T.}~\bibnamefont
  {Liebsch}}, \bibinfo {author} {\bibfnamefont {O.}~\bibnamefont {Plotzke}},
  \bibinfo {author} {\bibfnamefont {F.}~\bibnamefont {Heiser}}, \bibinfo
  {author} {\bibfnamefont {U.}~\bibnamefont {Hergenhahn}}, \bibinfo {author}
  {\bibfnamefont {O.}~\bibnamefont {Hemmers}}, \bibinfo {author} {\bibfnamefont
  {R.}~\bibnamefont {Wehlitz}}, \bibinfo {author} {\bibfnamefont
  {J.}~\bibnamefont {Viefhaus}}, \bibinfo {author} {\bibfnamefont
  {B.}~\bibnamefont {Langer}}, \bibinfo {author} {\bibfnamefont {S.~B.}\
  \bibnamefont {Whitfield}}, \ and\ \bibinfo {author} {\bibfnamefont
  {U.}~\bibnamefont {Becker}},\ }\href {\doibase 10.1103/PhysRevA.52.457}
  {\bibfield  {journal} {\bibinfo  {journal} {Phys. Rev. A}\ }\textbf {\bibinfo
  {volume} {52}},\ \bibinfo {pages} {457} (\bibinfo {year} {1995})}\BibitemShut
  {NoStop}%
\bibitem [{\citenamefont {Venuti}\ \emph {et~al.}(1999)\citenamefont {Venuti},
  \citenamefont {Stener}, \citenamefont {Alti},\ and\ \citenamefont
  {Decleva}}]{Venuti1999}%
  \BibitemOpen
  \bibfield  {author} {\bibinfo {author} {\bibfnamefont {M.}~\bibnamefont
  {Venuti}}, \bibinfo {author} {\bibfnamefont {M.}~\bibnamefont {Stener}},
  \bibinfo {author} {\bibfnamefont {G.~D.}\ \bibnamefont {Alti}}, \ and\
  \bibinfo {author} {\bibfnamefont {P.}~\bibnamefont {Decleva}},\ }\href
  {\doibase 10.1063/1.479220} {\bibfield  {journal} {\bibinfo  {journal} {The
  Journal of Chemical Physics}\ }\textbf {\bibinfo {volume} {111}},\ \bibinfo
  {pages} {4589} (\bibinfo {year} {1999})}\BibitemShut {NoStop}%
\bibitem [{\citenamefont {Decleva}\ \emph {et~al.}(2001)\citenamefont
  {Decleva}, \citenamefont {Furlan}, \citenamefont {Fronzoni},\ and\
  \citenamefont {Stener}}]{Decleva2001}%
  \BibitemOpen
  \bibfield  {author} {\bibinfo {author} {\bibfnamefont {P.}~\bibnamefont
  {Decleva}}, \bibinfo {author} {\bibfnamefont {S.}~\bibnamefont {Furlan}},
  \bibinfo {author} {\bibfnamefont {G.}~\bibnamefont {Fronzoni}}, \ and\
  \bibinfo {author} {\bibfnamefont {M.}~\bibnamefont {Stener}},\ }\href
  {\doibase 10.1016/s0009-2614(01)01166-6} {\bibfield  {journal} {\bibinfo
  {journal} {Chemical Physics Letters}\ }\textbf {\bibinfo {volume} {348}},\
  \bibinfo {pages} {363} (\bibinfo {year} {2001})}\BibitemShut {NoStop}%
\bibitem [{\citenamefont {Ton-That}\ \emph {et~al.}(2003)\citenamefont
  {Ton-That}, \citenamefont {Shard}, \citenamefont {Egger}, \citenamefont
  {Dhanak},\ and\ \citenamefont {Welland}}]{Ton-That2003}%
  \BibitemOpen
  \bibfield  {author} {\bibinfo {author} {\bibfnamefont {C.}~\bibnamefont
  {Ton-That}}, \bibinfo {author} {\bibfnamefont {A.~G.}\ \bibnamefont {Shard}},
  \bibinfo {author} {\bibfnamefont {S.}~\bibnamefont {Egger}}, \bibinfo
  {author} {\bibfnamefont {V.~R.}\ \bibnamefont {Dhanak}}, \ and\ \bibinfo
  {author} {\bibfnamefont {M.~E.}\ \bibnamefont {Welland}},\ }\href {\doibase
  10.1103/PhysRevB.67.155415} {\bibfield  {journal} {\bibinfo  {journal} {Phys.
  Rev. B}\ }\textbf {\bibinfo {volume} {67}},\ \bibinfo {pages} {155415}
  (\bibinfo {year} {2003})}\BibitemShut {NoStop}%
\bibitem [{\citenamefont {Li}\ \emph {et~al.}(2007)\citenamefont {Li},
  \citenamefont {Ni}, \citenamefont {Ying}, \citenamefont {Wang}, \citenamefont
  {Kurash},\ and\ \citenamefont {Qian}}]{Li2007}%
  \BibitemOpen
  \bibfield  {author} {\bibinfo {author} {\bibfnamefont {H.-N.}\ \bibnamefont
  {Li}}, \bibinfo {author} {\bibfnamefont {J.-F.}\ \bibnamefont {Ni}}, \bibinfo
  {author} {\bibfnamefont {L.-J.}\ \bibnamefont {Ying}}, \bibinfo {author}
  {\bibfnamefont {X.-X.}\ \bibnamefont {Wang}}, \bibinfo {author}
  {\bibfnamefont {I.}~\bibnamefont {Kurash}}, \ and\ \bibinfo {author}
  {\bibfnamefont {H.-J.}\ \bibnamefont {Qian}},\ }\href {\doibase
  10.1088/0953-8984/19/43/436223} {\bibfield  {journal} {\bibinfo  {journal}
  {Journal of Physics: Condensed Matter}\ }\textbf {\bibinfo {volume} {19}},\
  \bibinfo {pages} {436223} (\bibinfo {year} {2007})}\BibitemShut {NoStop}%
\bibitem [{\citenamefont {Daimon}\ \emph {et~al.}(1995)\citenamefont {Daimon},
  \citenamefont {Imada}, \citenamefont {Nishimoto},\ and\ \citenamefont
  {Suga}}]{Daimon1995}%
  \BibitemOpen
  \bibfield  {author} {\bibinfo {author} {\bibfnamefont {H.}~\bibnamefont
  {Daimon}}, \bibinfo {author} {\bibfnamefont {S.}~\bibnamefont {Imada}},
  \bibinfo {author} {\bibfnamefont {H.}~\bibnamefont {Nishimoto}}, \ and\
  \bibinfo {author} {\bibfnamefont {S.}~\bibnamefont {Suga}},\ }\href {\doibase
  10.1016/0368-2048(95)02478-6} {\bibfield  {journal} {\bibinfo  {journal}
  {Journal of Electron Spectroscopy and Related Phenomena}\ }\textbf {\bibinfo
  {volume} {76}},\ \bibinfo {pages} {487} (\bibinfo {year} {1995})}\BibitemShut
  {NoStop}%
\bibitem [{\citenamefont {Shirley}\ \emph {et~al.}(1995)\citenamefont
  {Shirley}, \citenamefont {Terminello}, \citenamefont {Santoni},\ and\
  \citenamefont {Himpsel}}]{Shirley1995}%
  \BibitemOpen
  \bibfield  {author} {\bibinfo {author} {\bibfnamefont {E.~L.}\ \bibnamefont
  {Shirley}}, \bibinfo {author} {\bibfnamefont {L.~J.}\ \bibnamefont
  {Terminello}}, \bibinfo {author} {\bibfnamefont {A.}~\bibnamefont {Santoni}},
  \ and\ \bibinfo {author} {\bibfnamefont {F.~J.}\ \bibnamefont {Himpsel}},\
  }\href {http://link.aps.org/abstract/PRB/v51/p13614} {\bibfield  {journal}
  {\bibinfo  {journal} {Physical Review B}\ }\textbf {\bibinfo {volume} {51}},\
  \bibinfo {pages} {13614} (\bibinfo {year} {1995})}\BibitemShut {NoStop}%
\bibitem [{\citenamefont {Martins}\ \emph {et~al.}(1991)\citenamefont
  {Martins}, \citenamefont {Troullier},\ and\ \citenamefont
  {Weaver}}]{Martins1991}%
  \BibitemOpen
  \bibfield  {author} {\bibinfo {author} {\bibfnamefont {J.~L.}\ \bibnamefont
  {Martins}}, \bibinfo {author} {\bibfnamefont {N.}~\bibnamefont {Troullier}},
  \ and\ \bibinfo {author} {\bibfnamefont {J.}~\bibnamefont {Weaver}},\ }\href
  {\doibase 10.1016/0009-2614(91)85149-q} {\bibfield  {journal} {\bibinfo
  {journal} {Chemical Physics Letters}\ }\textbf {\bibinfo {volume} {180}},\
  \bibinfo {pages} {457} (\bibinfo {year} {1991})}\BibitemShut {NoStop}%
\bibitem [{\citenamefont {Weaver}\ and\ \citenamefont
  {Poirier}(1994)}]{Weaver1994}%
  \BibitemOpen
  \bibfield  {author} {\bibinfo {author} {\bibfnamefont {J.~H.}\ \bibnamefont
  {Weaver}}\ and\ \bibinfo {author} {\bibfnamefont {D.~M.}\ \bibnamefont
  {Poirier}},\ }\href
  {https://www.ebook.de/de/product/15173315/fullerenes.html} {\emph {\bibinfo
  {title} {Solid State Physics}}},\ edited by\ \bibinfo {editor} {\bibfnamefont
  {H.}~\bibnamefont {Ehrenreich}}\ and\ \bibinfo {editor} {\bibfnamefont
  {F.}~\bibnamefont {Spaepen}},\ \bibinfo {series} {Advances in Research and
  Applications}, Vol.~\bibinfo {volume} {48}\ (\bibinfo  {publisher} {Elsevier
  S\&T},\ \bibinfo {year} {1994})\BibitemShut {NoStop}%
\bibitem [{\citenamefont {Kuzmany}(1998)}]{Kuzmany1998}%
  \BibitemOpen
  \bibfield  {author} {\bibinfo {author} {\bibfnamefont {H.}~\bibnamefont
  {Kuzmany}},\ }\href {\doibase 10.1002/piuz.19980290104} {\bibfield  {journal}
  {\bibinfo  {journal} {Physik in unserer Zeit}\ }\textbf {\bibinfo {volume}
  {29}},\ \bibinfo {pages} {16} (\bibinfo {year} {1998})}\BibitemShut {NoStop}%
\bibitem [{\citenamefont {He}\ \emph {et~al.}(1995)\citenamefont {He},
  \citenamefont {Bao}, \citenamefont {Yu},\ and\ \citenamefont {Xu}}]{He1995}%
  \BibitemOpen
  \bibfield  {author} {\bibinfo {author} {\bibfnamefont {P.}~\bibnamefont
  {He}}, \bibinfo {author} {\bibfnamefont {S.}~\bibnamefont {Bao}}, \bibinfo
  {author} {\bibfnamefont {C.}~\bibnamefont {Yu}}, \ and\ \bibinfo {author}
  {\bibfnamefont {Y.}~\bibnamefont {Xu}},\ }\href
  {http://www.sciencedirect.com/science/article/pii/0039602895000364}
  {\bibfield  {journal} {\bibinfo  {journal} {Surface Science}\ }\textbf
  {\bibinfo {volume} {328}},\ \bibinfo {pages} {287} (\bibinfo {year}
  {1995})}\BibitemShut {NoStop}%
\bibitem [{\citenamefont {Zhang}\ \emph {et~al.}(2011)\citenamefont {Zhang},
  \citenamefont {Richard}, \citenamefont {Qian}, \citenamefont {Xu},
  \citenamefont {Dai},\ and\ \citenamefont {Ding}}]{Zhang2011}%
  \BibitemOpen
  \bibfield  {author} {\bibinfo {author} {\bibfnamefont {P.}~\bibnamefont
  {Zhang}}, \bibinfo {author} {\bibfnamefont {P.}~\bibnamefont {Richard}},
  \bibinfo {author} {\bibfnamefont {T.}~\bibnamefont {Qian}}, \bibinfo {author}
  {\bibfnamefont {Y.-M.}\ \bibnamefont {Xu}}, \bibinfo {author} {\bibfnamefont
  {X.}~\bibnamefont {Dai}}, \ and\ \bibinfo {author} {\bibfnamefont
  {H.}~\bibnamefont {Ding}},\ }\href {\doibase 10.1063/1.3585113} {\bibfield
  {journal} {\bibinfo  {journal} {Review of Scientific Instruments}\ }\textbf
  {\bibinfo {volume} {82}},\ \bibinfo {pages} {043712} (\bibinfo {year}
  {2011})}\BibitemShut {NoStop}%
\bibitem [{\citenamefont {Troullier}\ and\ \citenamefont
  {Martins}(1992)}]{Troullier1992}%
  \BibitemOpen
  \bibfield  {author} {\bibinfo {author} {\bibfnamefont {N.}~\bibnamefont
  {Troullier}}\ and\ \bibinfo {author} {\bibfnamefont {J.~L.}\ \bibnamefont
  {Martins}},\ }\href {\doibase 10.1103/PhysRevB.46.1754} {\bibfield  {journal}
  {\bibinfo  {journal} {Phys. Rev. B}\ }\textbf {\bibinfo {volume} {46}},\
  \bibinfo {pages} {1754} (\bibinfo {year} {1992})}\BibitemShut {NoStop}%
\bibitem [{\citenamefont {Haddon}(1992)}]{Haddon1992}%
  \BibitemOpen
  \bibfield  {author} {\bibinfo {author} {\bibfnamefont {R.~C.}\ \bibnamefont
  {Haddon}},\ }\href {\doibase 10.1021/ar00015a005} {\bibfield  {journal}
  {\bibinfo  {journal} {Accounts of Chemical Research}\ }\textbf {\bibinfo
  {volume} {25}},\ \bibinfo {pages} {127} (\bibinfo {year} {1992})}\BibitemShut
  {NoStop}%
\bibitem [{\citenamefont {Cooper}\ and\ \citenamefont
  {Manson}(1969)}]{Cooper1969}%
  \BibitemOpen
  \bibfield  {author} {\bibinfo {author} {\bibfnamefont {J.~W.}\ \bibnamefont
  {Cooper}}\ and\ \bibinfo {author} {\bibfnamefont {S.~T.}\ \bibnamefont
  {Manson}},\ }\href {\doibase 10.1103/PhysRev.177.157} {\bibfield  {journal}
  {\bibinfo  {journal} {Phys. Rev.}\ }\textbf {\bibinfo {volume} {177}},\
  \bibinfo {pages} {157} (\bibinfo {year} {1969})}\BibitemShut {NoStop}%
\bibitem [{\citenamefont {He}\ \emph {et~al.}(2007)\citenamefont {He},
  \citenamefont {Arita}, \citenamefont {Namatame}, \citenamefont {Taniguchi},
  \citenamefont {Li},\ and\ \citenamefont {Li}}]{He2007}%
  \BibitemOpen
  \bibfield  {author} {\bibinfo {author} {\bibfnamefont {S.}~\bibnamefont
  {He}}, \bibinfo {author} {\bibfnamefont {M.}~\bibnamefont {Arita}}, \bibinfo
  {author} {\bibfnamefont {H.}~\bibnamefont {Namatame}}, \bibinfo {author}
  {\bibfnamefont {M.}~\bibnamefont {Taniguchi}}, \bibinfo {author}
  {\bibfnamefont {H.-N.}\ \bibnamefont {Li}}, \ and\ \bibinfo {author}
  {\bibfnamefont {H.-Y.}\ \bibnamefont {Li}},\ }\href
  {http://stacks.iop.org/0953-8984/19/i=2/a=026202} {\bibfield  {journal}
  {\bibinfo  {journal} {Journal of Physics: Condensed Matter}\ }\textbf
  {\bibinfo {volume} {19}},\ \bibinfo {pages} {026202} (\bibinfo {year}
  {2007})}\BibitemShut {NoStop}%
\bibitem [{\citenamefont {Cooper}\ and\ \citenamefont
  {Zare}(1968)}]{Cooper1968}%
  \BibitemOpen
  \bibfield  {author} {\bibinfo {author} {\bibfnamefont {J.}~\bibnamefont
  {Cooper}}\ and\ \bibinfo {author} {\bibfnamefont {R.~N.}\ \bibnamefont
  {Zare}},\ }\href {\doibase 10.1063/1.1668742} {\bibfield  {journal} {\bibinfo
   {journal} {The Journal of Chemical Physics}\ }\textbf {\bibinfo {volume}
  {48}},\ \bibinfo {pages} {942} (\bibinfo {year} {1968})}\BibitemShut
  {NoStop}%
\bibitem [{\citenamefont {Westphal}\ \emph {et~al.}(1990)\citenamefont
  {Westphal}, \citenamefont {Bansmann}, \citenamefont {Getzlaff},\ and\
  \citenamefont {Sch{\"o}nhense}}]{Westphal1990}%
  \BibitemOpen
  \bibfield  {author} {\bibinfo {author} {\bibfnamefont {C.}~\bibnamefont
  {Westphal}}, \bibinfo {author} {\bibfnamefont {J.}~\bibnamefont {Bansmann}},
  \bibinfo {author} {\bibfnamefont {M.}~\bibnamefont {Getzlaff}}, \ and\
  \bibinfo {author} {\bibfnamefont {G.}~\bibnamefont {Sch{\"o}nhense}},\ }\href
  {\doibase 10.1016/0042-207x(90)90281-3} {\bibfield  {journal} {\bibinfo
  {journal} {Vacuum}\ }\textbf {\bibinfo {volume} {41}},\ \bibinfo {pages} {87}
  (\bibinfo {year} {1990})}\BibitemShut {NoStop}%
\bibitem [{\citenamefont {Sch{\"o}nhense}(1990)}]{Schoenhense1990}%
  \BibitemOpen
  \bibfield  {author} {\bibinfo {author} {\bibfnamefont {G.}~\bibnamefont
  {Sch{\"o}nhense}},\ }\href {\doibase 10.1088/0031-8949/1990/t31/035}
  {\bibfield  {journal} {\bibinfo  {journal} {Physica Scripta}\ }\textbf
  {\bibinfo {volume} {T31}},\ \bibinfo {pages} {255} (\bibinfo {year}
  {1990})}\BibitemShut {NoStop}%
\bibitem [{\citenamefont {Sch{\"o}nhense}\ \emph {et~al.}(1992)\citenamefont
  {Sch{\"o}nhense}, \citenamefont {Westphal}, \citenamefont {Bansmann},\ and\
  \citenamefont {Getzlaff}}]{Schoenhense1992}%
  \BibitemOpen
  \bibfield  {author} {\bibinfo {author} {\bibfnamefont {G.}~\bibnamefont
  {Sch{\"o}nhense}}, \bibinfo {author} {\bibfnamefont {C.}~\bibnamefont
  {Westphal}}, \bibinfo {author} {\bibfnamefont {J.}~\bibnamefont {Bansmann}},
  \ and\ \bibinfo {author} {\bibfnamefont {M.}~\bibnamefont {Getzlaff}},\
  }\href {\doibase 10.1209/0295-5075/17/8/011} {\bibfield  {journal} {\bibinfo
  {journal} {Europhysics Letters ({EPL})}\ }\textbf {\bibinfo {volume} {17}},\
  \bibinfo {pages} {727} (\bibinfo {year} {1992})}\BibitemShut {NoStop}%
\bibitem [{\citenamefont {Dubs}\ \emph {et~al.}(1985)\citenamefont {Dubs},
  \citenamefont {Dixit},\ and\ \citenamefont {McKoy}}]{Dubs1985}%
  \BibitemOpen
  \bibfield  {author} {\bibinfo {author} {\bibfnamefont {R.~L.}\ \bibnamefont
  {Dubs}}, \bibinfo {author} {\bibfnamefont {S.~N.}\ \bibnamefont {Dixit}}, \
  and\ \bibinfo {author} {\bibfnamefont {V.}~\bibnamefont {McKoy}},\ }\href
  {\doibase 10.1103/physrevb.32.8389} {\bibfield  {journal} {\bibinfo
  {journal} {Physical Review B}\ }\textbf {\bibinfo {volume} {32}},\ \bibinfo
  {pages} {8389} (\bibinfo {year} {1985})}\BibitemShut {NoStop}%
\bibitem [{\citenamefont {Hasegawa}\ \emph
  {et~al.}(1998{\natexlab{b}})\citenamefont {Hasegawa}, \citenamefont
  {Miyamae}, \citenamefont {Yakushi}, \citenamefont {Inokuchi}, \citenamefont
  {Seki},\ and\ \citenamefont {Ueno}}]{Hasegawa1998}%
  \BibitemOpen
  \bibfield  {author} {\bibinfo {author} {\bibfnamefont {S.}~\bibnamefont
  {Hasegawa}}, \bibinfo {author} {\bibfnamefont {T.}~\bibnamefont {Miyamae}},
  \bibinfo {author} {\bibfnamefont {K.}~\bibnamefont {Yakushi}}, \bibinfo
  {author} {\bibfnamefont {H.}~\bibnamefont {Inokuchi}}, \bibinfo {author}
  {\bibfnamefont {K.}~\bibnamefont {Seki}}, \ and\ \bibinfo {author}
  {\bibfnamefont {N.}~\bibnamefont {Ueno}},\ }\href {\doibase
  http://dx.doi.org/10.1016/S0368-2048(97)00208-9} {\bibfield  {journal}
  {\bibinfo  {journal} {Journal of Electron Spectroscopy and Related
  Phenomena}\ }\textbf {\bibinfo {volume} {88}},\ \bibinfo {pages} {891 }
  (\bibinfo {year} {1998}{\natexlab{b}})}\BibitemShut {NoStop}%
\bibitem [{\citenamefont {Satpathy}\ \emph {et~al.}(1992)\citenamefont
  {Satpathy}, \citenamefont {Antropov}, \citenamefont {Andersen}, \citenamefont
  {Jepsen}, \citenamefont {Gunnarsson},\ and\ \citenamefont
  {Liechtenstein}}]{Satpathy1992}%
  \BibitemOpen
  \bibfield  {author} {\bibinfo {author} {\bibfnamefont {S.}~\bibnamefont
  {Satpathy}}, \bibinfo {author} {\bibfnamefont {V.~P.}\ \bibnamefont
  {Antropov}}, \bibinfo {author} {\bibfnamefont {O.~K.}\ \bibnamefont
  {Andersen}}, \bibinfo {author} {\bibfnamefont {O.}~\bibnamefont {Jepsen}},
  \bibinfo {author} {\bibfnamefont {O.}~\bibnamefont {Gunnarsson}}, \ and\
  \bibinfo {author} {\bibfnamefont {A.~I.}\ \bibnamefont {Liechtenstein}},\
  }\href {\doibase 10.1103/physrevb.46.1773} {\bibfield  {journal} {\bibinfo
  {journal} {Physical Review B}\ }\textbf {\bibinfo {volume} {46}},\ \bibinfo
  {pages} {1773} (\bibinfo {year} {1992})}\BibitemShut {NoStop}%
\bibitem [{Pho()}]{PhotENote}%
  \BibitemOpen
  \href@noop {} {}\bibinfo {note} {A measurement in a wider range of photon
  energy such as R{\"u}del et al.~\cite{Ruedel2002} should give a more precise
  value after FFT analysis.}\BibitemShut {Stop}%
\bibitem [{\citenamefont {{Braga}}\ \emph {et~al.}(1991)\citenamefont
  {{Braga}}, \citenamefont {{Larsson}}, \citenamefont {{Rosen}},\ and\
  \citenamefont {{Volosov}}}]{Braga1991}%
  \BibitemOpen
  \bibfield  {author} {\bibinfo {author} {\bibfnamefont {M.}~\bibnamefont
  {{Braga}}}, \bibinfo {author} {\bibfnamefont {S.}~\bibnamefont {{Larsson}}},
  \bibinfo {author} {\bibfnamefont {A.}~\bibnamefont {{Rosen}}}, \ and\
  \bibinfo {author} {\bibfnamefont {A.}~\bibnamefont {{Volosov}}},\ }\href@noop
  {} {\bibfield  {journal} {\bibinfo  {journal} {Astronomy and Astrophysics}\
  }\textbf {\bibinfo {volume} {245}},\ \bibinfo {pages} {232} (\bibinfo {year}
  {1991})}\BibitemShut {NoStop}%
\end{thebibliography}%

\end{document}